\documentclass[12pt]{article}

\usepackage{amsmath}
\usepackage{graphicx}
\usepackage{enumerate}
\usepackage{natbib}
\usepackage{url} 

\newcommand{\blind}{1}

\usepackage{amsthm,amsmath,natbib,mathtools,amssymb}
\RequirePackage[colorlinks,citecolor=blue,urlcolor=blue]{hyperref}
\usepackage{graphicx}
\usepackage{subcaption}
\usepackage{algorithm}
\usepackage{algorithmic}
\usepackage{cleveref}
\usepackage{enumitem}
\usepackage{verbatim}

\numberwithin{equation}{section}
\theoremstyle{plain}
\newtheorem{thm}{Theorem}[section]
\newtheorem{defn}{Definition}[section]

\begin{document}

\def\spacingset#1{\renewcommand{\baselinestretch}%
{#1}\small\normalsize} \spacingset{1}


\if1\blind
{
  \title{\bf Dirichlet Depths for Point Process}
  \author{Kai Qi, Yang Chen, Wei Wu \\
    Department of Statistics, Florida State University\\
    Tallahassee, FL 32306-4330}
  \maketitle
} \fi

\if0\blind
{
  \bigskip
  \bigskip
  \bigskip
  \begin{center}
    {\LARGE\bf Dirichlet Depths for Point Process}
\end{center}
  \medskip
} \fi

\bigskip
\begin{abstract}
Statistical depths have been well studied for multivariate and functional data over the past few decades, but remain under-explored for point processes.  A first attempt on the notion of point process depth was conducted recently where the depth was defined as a weighted product of two terms: (1) the probability of the number of events in each process and (2) the depth of the event times conditioned on the number of events by using a Mahalanobis depth.  We point out that multivariate depths such as the Mahalanobis depth cannot be directly used because they often neglect the important ordered property in the point process events. To deal with this problem, we propose a model-based approach for point processes systematically.  In particular, we develop a Dirichlet-distribution-based framework on the conditional depth term, where the new methods are referred to as Dirichlet depths. We examine the mathematical properties of the new depths and conduct the asymptotic analysis.  In addition, we illustrate the new methods using various simulated and real experiment data.  It is found that the proposed framework provides a proper center-outward rank and the new methods have superior decoding performance to previous methods in two neural spike train datasets.  
\end{abstract}

\noindent%
{\it Keywords:}  Point process, Dirichlet depth, Poisson process, Time warping, Neural spike trains
\vfill

%
%
%
%
%
%
%
%
%


\newpage
\spacingset{1.5} 

\section{Introduction}

Point process models have been well studied for many decades and widely applied in various disciplines, such as geography, seismology, astronomy, neuroscience, and so on.  Those models are mainly focused on representing observations at each given time/location and have limited capability to measure the center-outward ranks of data. The center-outward rank, often referred to as statistical depth (depth for short), is a powerful tool to understand the features of underlying distribution such as spread and shape \citep{Liu99}. The study on depth has been focused on multivariate data and functional data \citep{Zuo2000,  Pintado09, Mosler12}.  In practice, depth has been successfully applied to address various practical problems such as classification \citep{Lange2014}, outlier detection \citep{Chen09}, and diagnostics of nonnormality \citep{Liu99}.

The notion of statistical depth was first introduced and systematically studied on multivariate data by \cite{Tukey75}.  Since then, various definitions of multivariate depth have been proposed such as the convex hull peeling depth \citep{Barnett76}, Oja depth \citep{Oja83}, simplicial depth \citep{Liu90}, Mahalanobis depth \citep{Liu93}, and likelihood depth \citep{Fraiman99}. As an axiomatic approach, a more general notion of depth for multivariate data was proposed by \cite{Zuo2000}, in which they summarized four desirable properties for multivariate depths, namely affine invariance, maximality at the center, monotonicity relative to the deepest points, and vanishing at infinity. In addition to multivariate data, depth for functional observations has received extensive attention in recent years \citep{Pintado09, Mosler12}.   Similar to the axiomatic approach in \citep{Zuo2000}, \cite{Nieto2016} provided a general definition of functional depth through six desirable properties, namely distance invariance, maximality at the center, decreasing with respect to the deepest point, upper semi-continuity in the function space, receptivity to convex hull with across the domain, and continuity in the probability measure.
Mathematical theories have also been extensively studied in majority of depth methods. For example, \cite{Nolan92} and \cite{masse2004} studied the convergence behavior of the halfspace depth and depth trimmed regions, and \cite{Koshevoy97} studied the convergence behavior of the Zonoid depth. Furthermore, \cite{Dyckerhoff16} discussed the connections between different types of convergence for multivariate depths. \cite{zuo2000b} studied the structural properties of trimmed regions, such as affine equivariance, nestedness, connectedness, and compactness.  

Our goal in this paper is to study the notion of statistical depth in temporal point process data.  This is an under-explored area. The only previous work is given in  \citep{Liu2017}, where the authors introduced the notion of depth in point process using a basic Mahalanobis depth.  Note that given the number of the events in a point process, the distribution of these events follow a multivariate framework.  However, we point out that the multivariate depths cannot be directly used for point process data.  This is because i) the number of events is a random variable, which is not described by the multivariate depths; ii) the events in a point process are an ordered sequence in a given (often finite) time domain.  To the best of our knowledge, none of the multivariate depths studied the center-outward rank on ordered data. 
 
Using mathematical notation, let $S$ denotes the set of all point processes in a time domain $[T_1,T_2]$.  Then an observed realization $s=(s_1, s_2, \cdots, s_k) \in S $ can be treated as a vector in $\mathbb{R}^k$, where $|s|=k$ is the cardinality of $s$. This cardinality $k$ can be any nonnegative integer.  By the nature of temporal point process, the events $(s_1, s_2, \cdots, s_k)$ are ordered in a natural way as $T_1 \leq s_1 < s_2 < \cdots < s_k \leq T_2$. Traditional depths defined on multivariate data neglect the importance of this order and will not be suitable for point process events. For example, suppose we inter-change the position of $s_2$ with $s_1$ and let $s'=(s_2, s_1, s_3, \cdots, s_k)$, traditional depth functions may still assign some positive depth value to $s'$, but $s'$ appears to be an outlier with zero probability, which is expected to have a zero depth value.

A depth function needs to take into account two types of randomness in a temporal point process $s$:
(1) the number of events, or cardinality, in the process, denoted as $|s|$, and
(2) the conditional distribution of these events given $|s|$. 
The notion of depth for point process was first studied by \cite{Liu2017}, where the authors defined a new depth framework as a weighted product of two terms: (1) the normalized probability of the number of events in each process and (2) the depth of the event times conditioned on the number of events by using the Mahalanobis depth. The weighted product is an appropriate way to address the two types of randomness.  However, the Gaussian-kernel-based Mahalanobis conditional depth neglect the bounded and ordered property of the events.  Here we use an example to illustrate how Gaussian-kernel-based conditional depth is inappropriate for the point process.  The detailed method is given in the Methods Section. 

\begin{figure}[h]
\centering
  \begin{subfigure}[b]{0.3\textwidth}
\includegraphics[width=\textwidth]{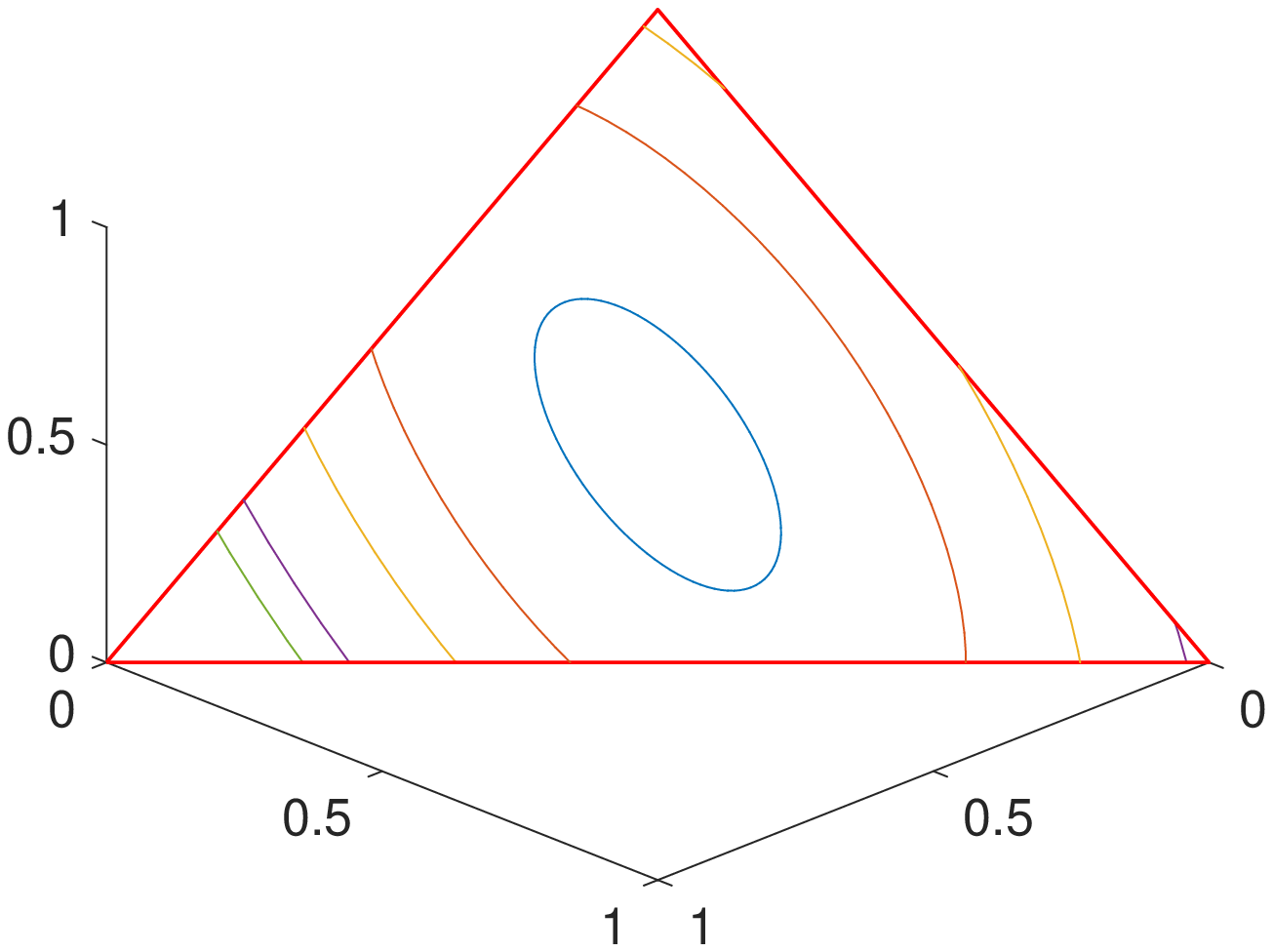}
\caption{Mahalanobis Depth}
\label{}
\end{subfigure}%
\begin{subfigure}[b]{0.3\textwidth}
\includegraphics[width=\textwidth]{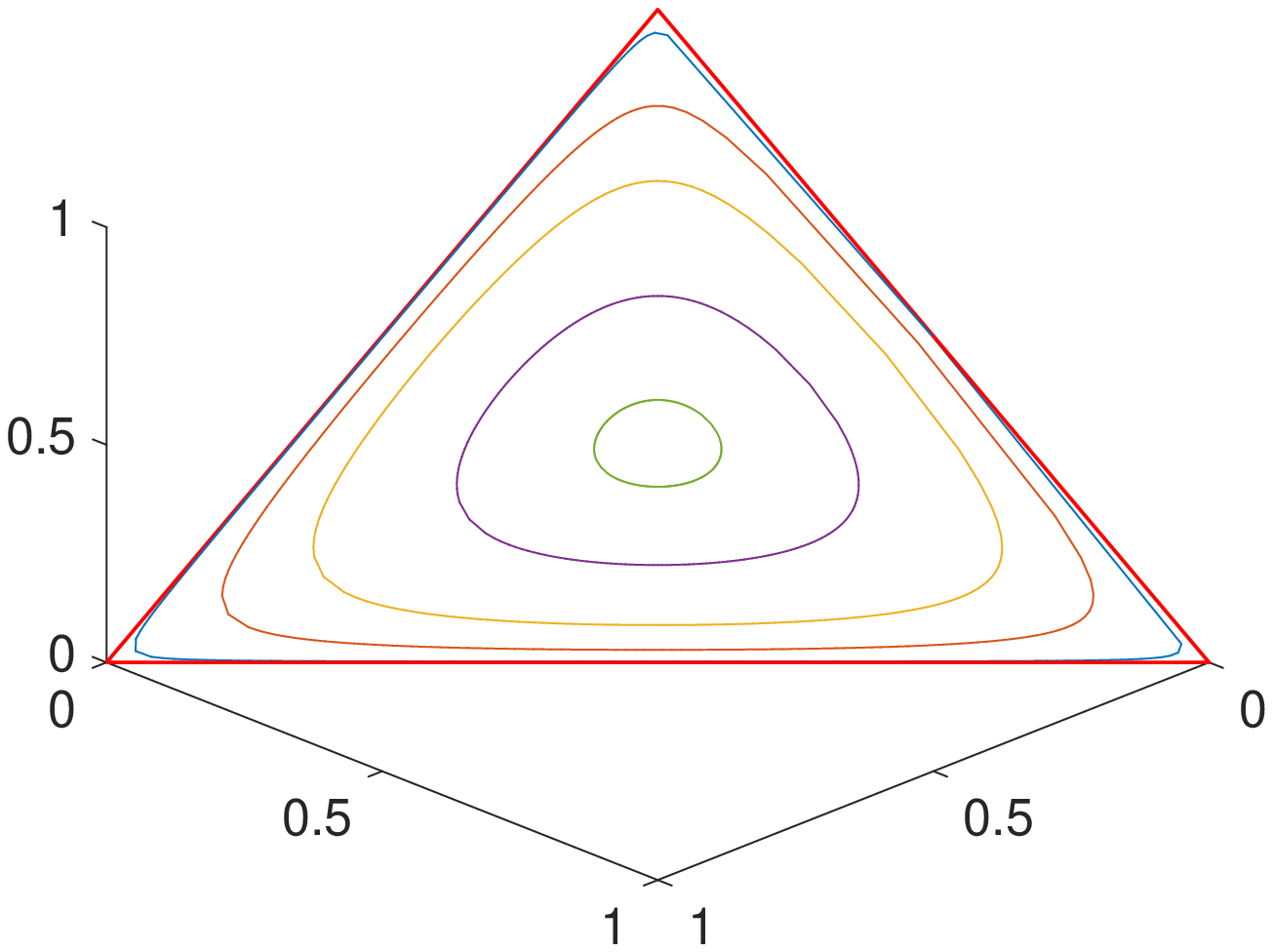}
\caption{Proposed Depth}
\label{}
\end{subfigure}%

\caption[]{Comparison of the Mahalanobis depth and the proposed depth. (a) A 2-dimensional simplex which indicates the inter-event times of a homogeneous Poisson process in $[0,1]$, conditional on the cardinality $|s|=2$.  The conditional depth contours are based on the Mahalanobis depth. (b) Same as (a) except using the proposed depth.  }
\label{sample-figure}
\end{figure}


Basically, for a homogeneous Poisson process that only has two ordered events $s_1$ and $s_2$ in the time interval $[0,1]$, the three inter-event times are: $s_1$, $s_2-s_1$ and $1-s_2$.  These three intervals are nonnegative with the sum being 1, and therefore form a 2-dimensional simplex (i.e. a triangle)  as shown in Figure \ref{sample-figure}. 
For the Mahalanobis depth, Gaussian kernel is applied on the events, and therefore the inter-event times will also be represented by a Gaussian model. Typical Mahalanobis depth contours on the inter-event times are elliptical, as shown in Figure \ref{sample-figure}(a). We can see that such contours are not appropriate for the center-outward tendency since (1) the elliptical contours  do not match the triangular domain, and (2) the points on the border of simplex will be still assigned positive depth values by the Mahalanobis depth.

A more reasonable contour plot is shown in Figure \ref{sample-figure}(b), where all depth contours are triangle-like. Many non-parametric depths for multivariate data, such as the halfspace depth and convex hull peeling depth, could generate similar triangle-like shape contour if the sample size is large enough.  
However, a significant drawback of those methods is that the computational efficiency.  This is particularly an issue in high dimensional case, which is common to point process data. 
To address this issue, we focus on efficient parametric depths for point processes in this study. 

Defining center-outward ranks for point process observations is a timely and important research topic. The goal of this paper is to develop a new depth framework for point processes systematically. Based on the approach in \cite{Liu2017}, our proposed framework of depth function for point processes is also defined as a weighted product of two terms aforementioned. In this paper, we focus on introducing new conditional depth functions based on the Dirichlet distribution.  We will then discuss the desired mathematical properties and asymptotic behavior. 

The rest of this paper is organized as follows: In Sec. \ref{sec:methods}, we elaborate on the definitions of the new depths and provide computational procedures to effectively estimate them. The properties of the proposed depths are discussed in this section as well, followed by a thorough analysis with simulations. We then study the asymptotics of the sample Dirichlet depths in Sec. \ref{sec:asymptotic}.   In Sec \ref{sec:real}, we apply the new depths to decoding problems in two neural spike train datasets. Finally, we discuss and summarize the work in Sec. \ref{sec:summary}.  

\section{Methods}
\label{sec:methods}

In this section, we will at first review basic notation and then propose our new conditional depths for temporal point process. Since the new conditional depth functions are based on the Dirichlet distribution, we refer to them as the Dirichlet depths. 

\subsection{Notation and Depth Definition}

Let $S$ denote the set of all point processes in the time domain $[T_1,T_2]$.  For any non-negative integer $k$, let $S_k = \{s \in S \mid |s| = k\}= \{(s_1, \cdots, s_k) \in \mathbb R^k | T_1 \le s_1 \le \cdots \le s_k \le T_2\}$ denote the set of all point processes in $S$ with cardinality $|s|=k$.  Hence, $S = \bigcup_{k=0}^{\infty}S_k$.  
For any $s\in S$,  a depth function for point process is a map $D: S\rightarrow \mathbb R^+$ (set of nonnegative real numbers), $s\rightarrow D(s)$. 

As we have emphasized in Introduction, there are two types of randomness in a point process: (1) the number of events in each process, and (2) the conditional distribution of these event times. In \citep{Liu2017}, the number of events is modeled by a normalized Poisson mass function and the event times are modeled by a multivariate Gaussian distribution. The depth framework of a point process $s$ is then defined as a weighted product of two terms -- the normalized probability of having $|s|$ events and the conditional depth using the Mahalanobis depth. 
In this paper, we generalize this framework  (not limited to Poisson and Gaussian) by defining the depth as a weighted product of the following two terms:
(1) the normalized probability of the number of events in each process, and 
(2) the center-outward ranks on the event times conditional on the number of events.
The formal definition is given as follows:
\begin{defn}
Given a point process $s \in S$ on $[T_1,T_2]$, we define its depth $D(s)$ as:
\begin{equation}
D(s)=w(|s|)^rD_c (s \mid |s|)
\label{eq:depth}
\end{equation}
where $w(|s|)=\frac{P(|s|)}{\max_{k} P(k)}$ for $P(|s|)>0$ is the normalized probability based on the cardinality $|s|$, $r>0$ is the weight parameter, and $D_c (s \mid |s|)$ is the depth of $s$ conditioned on $|s|$. If $P(|s|)=0$, we define $w(|s|)=0$ and  $D(s)=0$. 
\end{defn}

The first term $w(|s|)$ only depends on the distribution of $|s|$, with $r$ as a tuning (weight) hyperparameter to balance its importance relative to the second term $D_c (s \mid |s|)$. As $r$ gets larger, $w(|s|)$ becomes a more dominant factor in the depth value $D(s)$. Various parametric or non-parametric methods can be adopted to estimate $w(|s|)$, and the choice of methods can depend on the goal of applications. In Sec. \ref{sec:real}, we adopt a mixture of Poisson probability mass functions to model $|s|$.  The parameter estimation can be done via a standard EM algorithm procedure.  

The second term $D_c (s \mid |s|)$ describes the conditional depth when the number of events $|s|$ is given. In principle, any multivariate depth can be used as the conditional depth for point process if we treat $s\in S_{|s|}$ as an $|s|$ dimensional vector. {\bf However, we point out that such an approach neglects two important conditions of point process on $[T_1,T_2]$: (1) the event times are constrained on $[T_1, T_2]$, and (2) there exists a natural order in the event time sequence. }
To address this issue, rather than defining conditional depth function on the original point process space, we propose to define conditional depth on inter-event times.

\subsection{Equivalent Representation and Desirable Properties}

The point processes we discussed are bounded and ordered, i.e. $T_1 \le s_1 \le s_2 \le \cdots \le s_k \le T_2$. Applying multivariate depth functions directly on $S_k$ as conditional depths will tend to neglect the boundedness and orderedness conditions. We propose to use inter-event times to represent a point process such that these important conditions  are naturally satisfied. 

\subsubsection{Representation using Inter-Event Times}  

It is well known that the point process can be equivalently represented by the inter-event times (IETs).  Here the IETs of a point process $s_1, s_2,\cdots,s_k$ on $[T_1, T_2]$ are given as $u_1 = s_1-T_1, u_2=s_2-s_1,\cdots,u_k = s_k - s_{k-1},u_{k+1}=T_2 - s_k$.  The IET sequence $(u_1,u_2,\cdots,u_{k+1})$ has $k$ degrees of freedom and in fact forms a $k$-dimensional simplex (scaled standard simplex) as:
$$X_k = \{u \in \mathbb{R}^{k+1} : u_1+u_2+\cdots +u_{k+1}=T_2-T_1, u_i \geq 0, i=1,2,\cdots,k+1\}. $$ 
This simplex $X_k$ is bounded by the {\it boundary} set 
$B_k = \{ u \in X_k: u_i = 0 \mbox{ for at least one } i \in \{1,2, \cdots, k+1\}\}.$
The points at boundary indicate a realization which has either two events happening simultaneously or one event happening at time $T_1$ or $T_2$.  Both situations indicate extreme realizations (often with zero probability density) of a point process.

Based on this IET representation, we look for a conditional depth defined on the $X_k$ simplex.  Notice that the normalized IET sequence $(\frac{u_1}{T_2-T_1}, \frac{u_2}{T_2-T_1}$, $\cdots$, $\frac{u_{k+1}}{T_2-T_1})$ has the constant sum of 1.  Therefore, one apparent option for the depth is the density function of Dirichlet distribution, which is commonly used as a prior in Bayesian statistics.  Here we review the Dirichlet distribution which will be used to derive our conditional depth function: The Dirichlet probability density function of order $m \ge 2$ with concentration parameter vector $\mathbf{a}=(a_1, a_2,\cdots, a_m)  \in \mathbb R^{m}$ with $a_i > 0$,  $i = 1, \cdots, m$, is given as:
\begin{equation}
f(x_1,x_2,\cdots,x_m ; a_1,a_2,\cdots,a_m)=\frac{\Gamma(\sum_{i=1}^m a_i)}{\prod_{i=1}^m \Gamma{(a_i)}} \prod_{i=1}^{m}x_{i}^{a_i-1} .
\label{eq:Diri}
\end{equation}
where $(x_1, x_2, \cdots, x_m)$ is in the standard $m-1$ simplex, i.e. $\sum_{i=1}^m x_i=1$ and $ x_i \ge 0, i=1,2,\cdots,m$. This density function is denoted as $Dirichlet(\mathbf a, m)$.

\subsubsection{Desirable properties of the conditional depth for Point Process} 
\label{subsection: discussion of properties}

In statistical depth literature, \cite{Zuo2000} and \cite{Nieto2016} proposed important and desirable properties for depth on multivariate and functional data, respectively. They further claim that a depth function should be defined through desirable properties. Motivated by this claim, we list and discuss five desirable properties for a conditional depth function for point process as follow.

\begin{itemize}
\item P-1, Continuity and vanishing at the boundary: \
Conditional depth for point process is a map from the simplex $X_k$ to $\mathbb{R}^+$. Since event times are continuous on the time domain, a minimal requirement for a proper conditional depth should be continuity. Also, an ideal conditional depth for point process should vanish at the boundary. 

\item P-2, Maximality at the center: \
This may be the most logical one among all properties since the center must have a maximal depth in a center-outward rank.  The notion of center can be defined using symmetric properties or the mathematical expectation.  


\item P-3, Monotonicity relative to the deepest point: \
This property is also intuitive as depth value should decrease from the center in a center-outward trend. 

\item P-4, Scale and shift invariance: \
The scale and shift invariance is a special case of the affine invariance in multivariate depth.  Basically, a good depth is expected to be invariant with respect to scaling and translation on the time domain. 



\item P-5, Time warping invariance: The variation of a point process must satisfy two conditions: 1) the events are in the domain $[T_1, T_2]$, and 2) the events remain the temporal order.  Such variation can be properly described by the set of time warping functions, defined as a boundary-preserving diffeomorphism  
 $\Gamma = \{ \gamma : [T_1, T_2] \rightarrow [T_1, T_2] \mid \gamma(T_1)=T_1, \gamma(T_2)=T_2, \dot \gamma > 0 \}$, where the dot indicates the first order derivative \citep{Anuj16}. 
 The time warping essentially allows any order-preserving nonlinear transformation of events in the given time domain. The time warping invariance also corresponds to the affine invariance in multivariate depths.


\end{itemize}
In the following sections, we will discuss all above properties in proposed conditional depths.

\subsection{Dirichlet Depth for Homogeneous Poisson Process}

We at first develop a Dirichlet depth for the most classical temporal point process -- homogeneous Poisson process (HPP). 

\subsubsection{Definition}

For an HPP, the first term $w(|s|)$ in Equation \eqref{eq:depth}  is simply the normalized Poisson probability on the number of events in the given process.  The challenge therefore stays on the conditional depth $D_c (s \mid |s|)$. 
As we have discussed, defining conditional depth for HPP on its IET representation will address the natural order issue, and ideally, the conditional depth should satisfy Properties P-1 to P-4. Property P-5 is not applicable here since any non-linear time warping on an HPP will make the process not homogeneous anymore. 


Before we step into the formal definition of Dirichlet depth, we first look at the connection between HPP and Dirichlet distribution. For an HPP, we have defined IETs $(u_i \ , \ i=1,\cdots,k+1)$ as mentioned earlier.  Conditioned on the number of events $k$, the normalized IETs $( \frac{u_i}{T_2-T_1} \ , \ i=1,2,\cdots,k+1)$ will satisfy two conditions: (1) They share the same support, a $k$-dimensional standard simplex, as the Dirichlet distribution (also true for any point process). (2) They follow a flat $Dirichlet(\{1,\cdots,1\},k+1)$ distribution, which is in fact a uniform distribution over the standard $k$-dimensional simplex. The detail proof is given in Part A of the Supplementary Materials. 

With a slight modification on Equation \eqref{eq:Diri}, we formally propose the Dirichlet depth for an HPP as follow:

\begin{defn}\label{defn:DirDepth HPP population}
Let $s=(s_1,s_2,\cdots,s_k)$ in $[T_1,T_2]$ be an observed homogeneous Poisson process.  Denote $s_0=T_1, \ s_{k+1}=T_2$.   The Dirichlet depth of $s$ (given $|s|$) is defined as:
\begin{equation}\label{eq:DirDepth HPP population}
D_c (s \mid |s|=k)=(k+1)\prod^{k+1}_{i=1}(\frac{s_i-s_{i-1}}{T_2-T_1})^{\frac{1}{k+1}}
\end{equation}
In particular, we have $D_c (s \mid |s|=0) = 1$.
\label{def:HPPdepth}
\end{defn}

The Dirichlet depth $D_c(s \mid |s|=k)$ for an HPP in Definition \ref{defn:DirDepth HPP population} describes the conditional depth of a realization when the number of events (cardinality) is known.
In Equation \eqref{eq:DirDepth HPP population}, we have set the concentration parameters of the Dirichlet distribution $a_i$ as $1+\frac{1}{k+1}$ for $\ i=1,2,\cdots,k+1$. This constant value makes the Dirichlet depth a concave function with maximum at the conditional mean (the derivation is given in the next subsection). The scale constant $(k+1)$ ensures $D_c(s \mid |s|=k)$ has an onto map to $[0, 1]$.  This normalization makes conditional depths comparable for observations across different number of events.


\subsubsection{Properties} 

When discussing the properties of a depth function on a given space (e.g. Dirichlet depth on a simplex), a central notion is the {\bf center} which depends on the underlying probability distribution.  
In this paper, conditioned on $|s|=k$ we take the center as the common mathematical expectation. That is, the center is
$$\theta_k= \mathbb{E}(s \mid |s|=k) = (\mathbb{E}(s_1 \mid |s| = k), \cdots, \mathbb{E}(s_k  \mid |s| = k)).$$ 
Using the derivation in Part A of the Supplementary Materials, the center has the following closed-form:  $$\theta_k = (\frac{T_2-T_1}{k+1},\frac{2(T_2-T_1)}{k+1},\cdots,\frac{k(T_2-T_1)}{k+1}).$$  Hence, 
the corresponding IET vector of conditional center is
$(\frac{T_2-T_1}{k+1},\frac{T_2-T_1}{k+1}$, $\cdots,\frac{T_2-T_1}{k+1})$.
On a $k$-dimensional simplex, this point is the same as the geometric center. 
For general point process other than the HPP, we will adopt the similar notion of center.

Now, we are ready to derive the important properties of the new conditional depth: 

{\bf P-1, Continuity and vanishing at the boundary:}
$D_c(s \mid |s|=k)$ is a continuous map from $S_k$ to $[0,1]$.  It is easy to verify that  $D_c(s \mid |s|=k) = 0$ if and only if the IET sequence of $s$ is in the boundary set $B_k$.

{\bf P-2, Maximality at the conditional mean (the center):}
Let $\theta_k=(\frac{T_2-T_1}{k+1},\frac{2(T_2-T_1)}{k+1}$, $\cdots,$ $\frac{k(T_2-T_1)}{k+1})$ denote the center of an HPP on $[T_1,T_2]$ with cardinality $k$.  Then $ D_c(\theta_k \mid |\theta_k|=k)= \sup_{s \in S_k} D_c(s \mid |s|=k ) = 1$.

{\bf P-3, Monotonicity relative to the deepest point:} 
For any $s \in S_k$ and $a \in [0, 1]$, we can prove that $ D_c(s \mid |s|=k)\leq D_c(\theta_k + a(s-\theta_k) \mid |s|=k)$. 

{\bf P-4, Scale and shift invariance:} 
For any scaling coefficient $a \in \mathbb{R^{+}}$ and translation $b \in \mathbb{R}$, we define a transformation of $s$ as $s'=a s + b \textbf{1}$ (where $\textbf{1} = (1, 1, \cdots, 1) \in \mathbb{R}^k$).  Let the time domain of $s$ be $[T_1, T_2]$. Then the time domain of $s'$ will be $[aT_1 +b, aT_2 +b]$.  If we include the time domain in the definition of Dirichlet depth, then we can show that $D_c(s \mid |s|=k, T_1, T_2)=D_c(s'\mid |s'|=k, aT_1+b, aT_2+b)$.

The proofs of all these properties are given in Part B of the Supplementary Materials.

\subsubsection{Illustration} 

In this subsection, we will at first examine the ranking performance of the Dirichlet depth on $S_2$.  We will then utilize the Dirichlet depth as the conditional depth in Equation \ref{eq:depth} to study the ranking performance on 100 HPP realizations.

\begin{figure}[h]
  \centering
  \begin{subfigure}[b]{0.4\textwidth}
    \includegraphics[width=\textwidth]{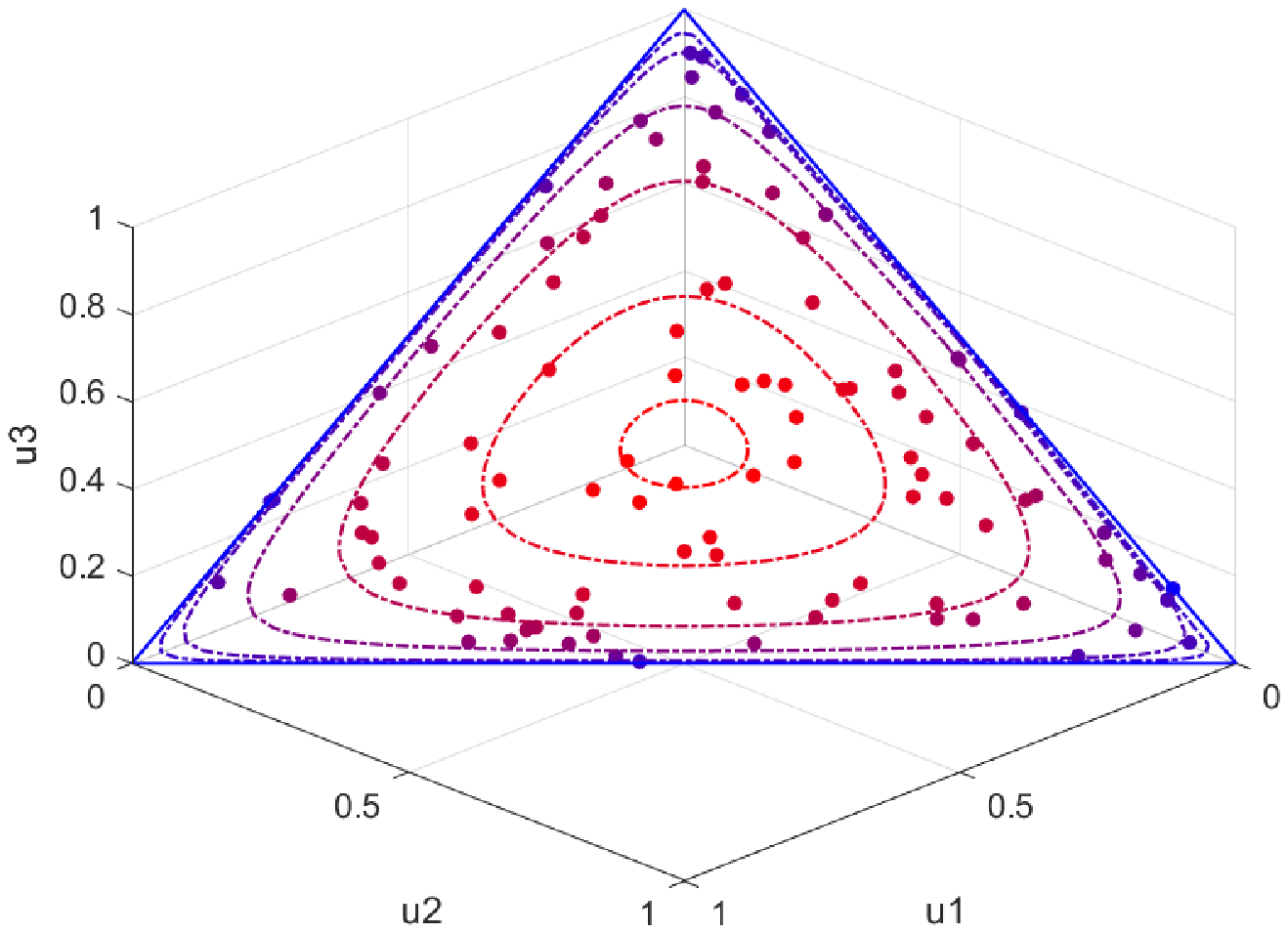}
    \caption{Dirichlet contours}
    \label{}
  \end{subfigure}%
  \begin{subfigure}[b]{0.4\textwidth}
    \includegraphics[width=\textwidth]{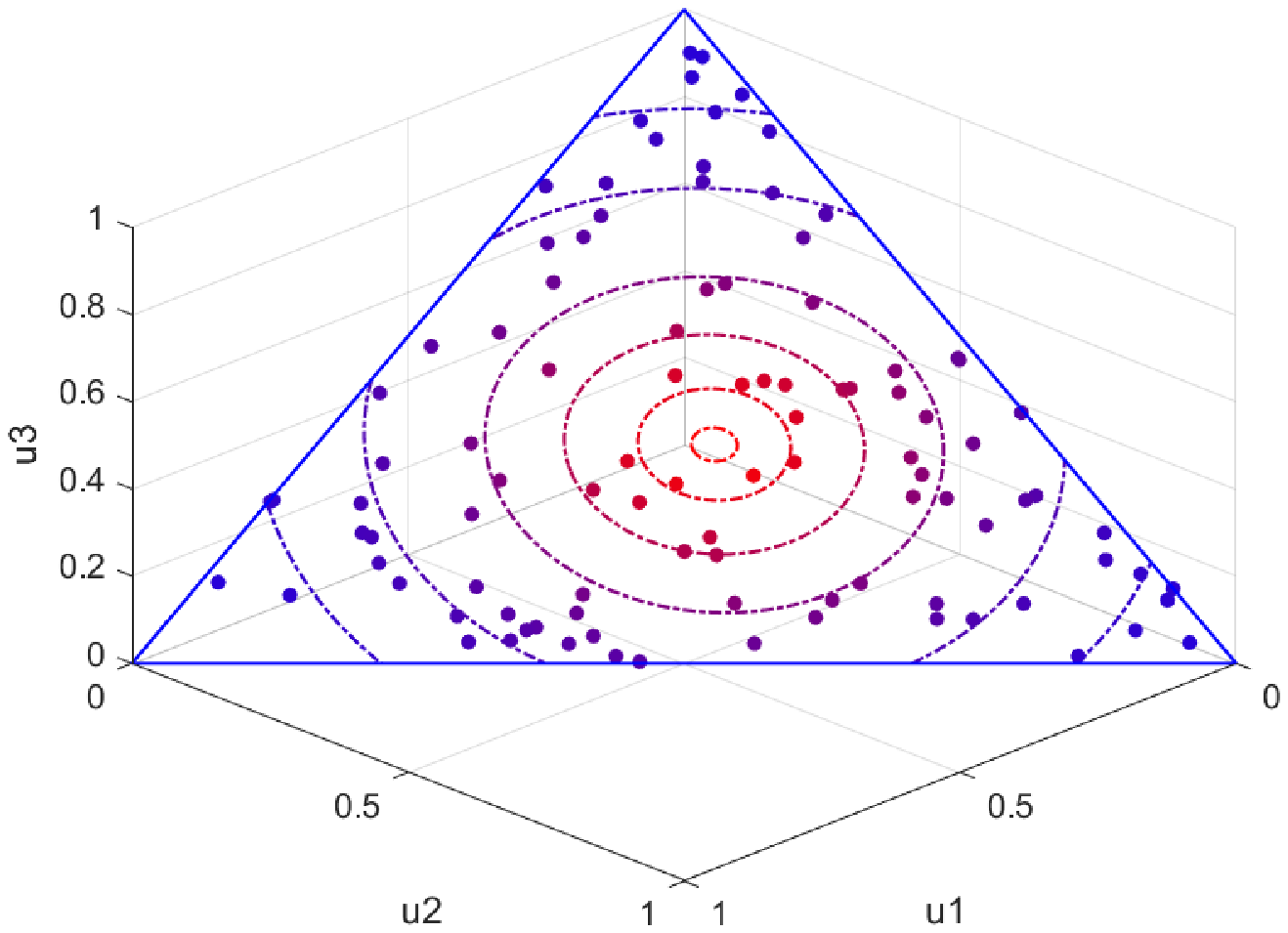}
    \caption{Mahalanobis contours}
    \label{}
  \end{subfigure}%

\caption[]{Example of Dirichlet depth and Mahalanobis depth for HPP conditioned on 2 events in $[0,1]$. The contours from outside to the center are with depth values 0, 0.2, 0.3, 0.5, 0.7, 0.9, and 0.99, respectively.  (a) Depth contours and IETs using the Dirichlet Depth. (b) Depth contours and IETs using the Mahalanobis Depth.}
\label{fig2}
\end{figure}

Conditioned on the cardinality $|s|=2$, the inter-event times are uniformly distributed on a 2-dimensional simplex. Here we simulate $100$ realizations from HPP conditional on $|s|=2$ in time interval $[0,1]$, and then apply both Dirichlet Depth for HPP (Equation \ref{eq:DirDepth HPP population}) and Mahalanobis depth for comparison. The result is shown in Figure \ref{fig2}.  We can see that compared with the (truncated) elliptic contours by the Mahalanobis depth, Dirichlet depth has smooth, triangle-like contours that are more compatible with the triangular IET domain. Also, the Mahalanobis depth assigns positive depth values for the points on the boundary, which is not reasonable in practice. 

\begin{figure}[h]
  \centering
  \begin{subfigure}[b]{0.35\textwidth}
    \includegraphics[width=\textwidth]{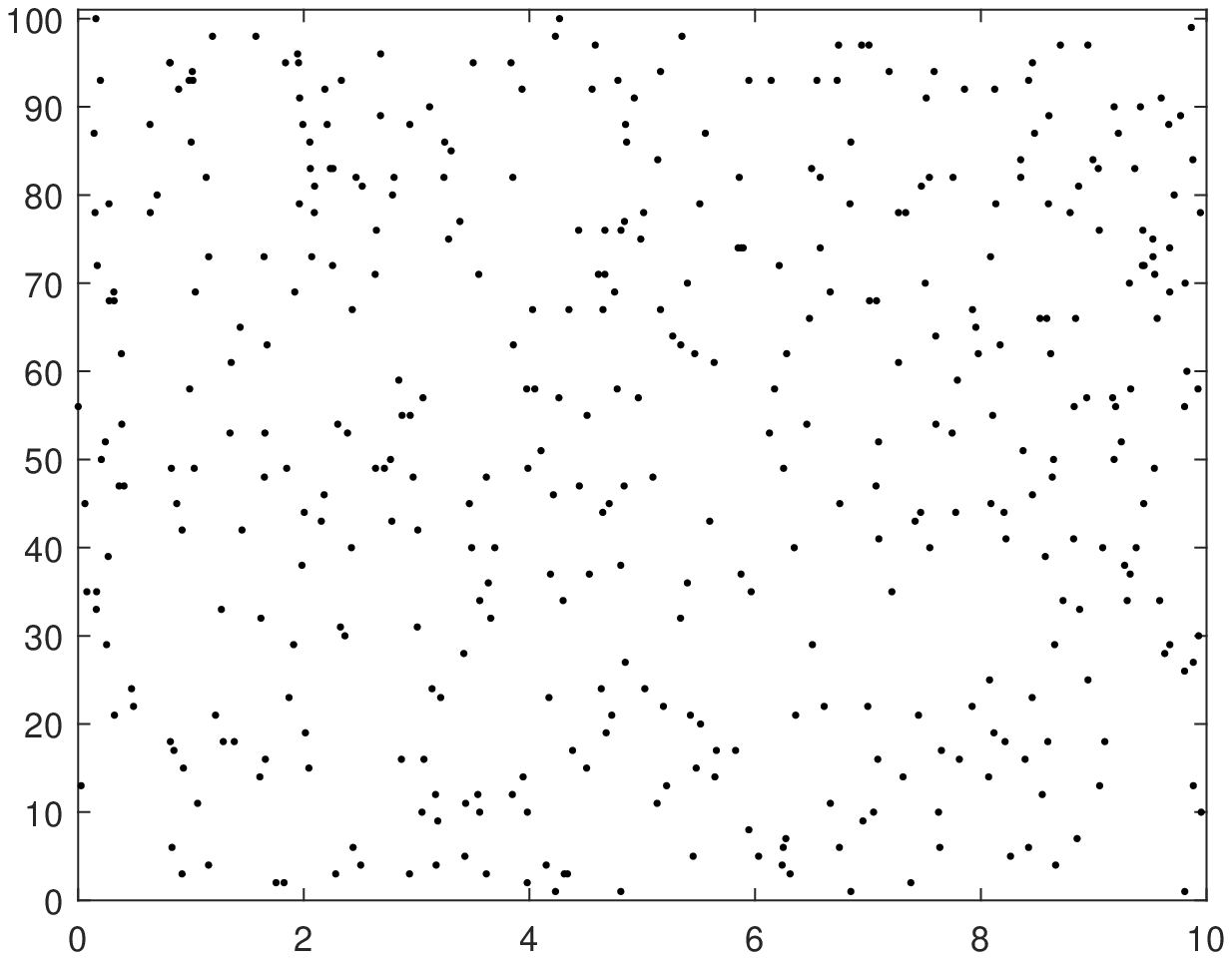}
    \caption{100 realizations}
    \label{}
  \end{subfigure}%
    \begin{subfigure}[b]{0.35\textwidth}
    \includegraphics[width=\textwidth]{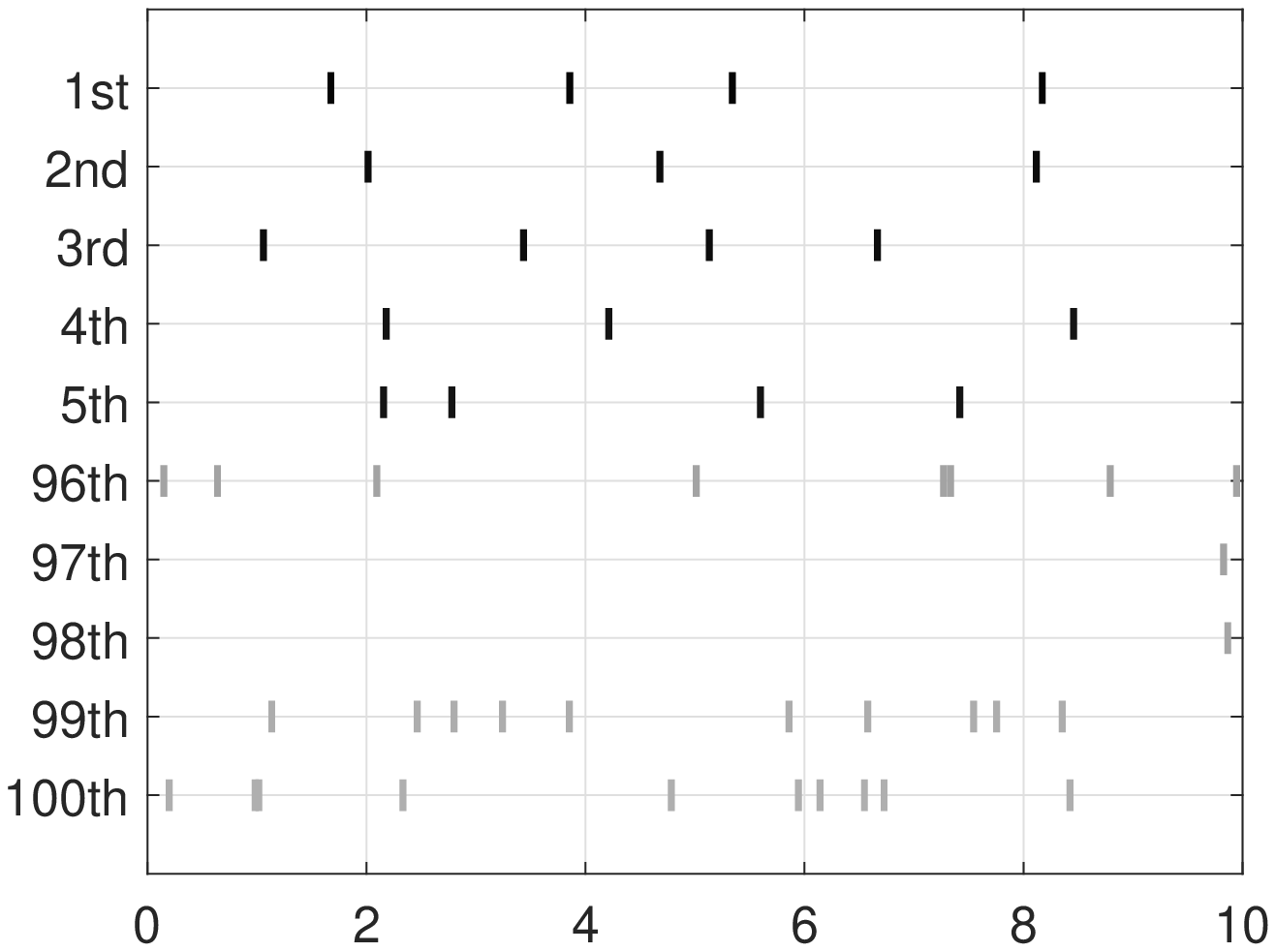}
    \caption{$r$=1}
    \label{HPPr1}
  \end{subfigure}%
  \begin{subfigure}[b]{0.35\textwidth}
    \includegraphics[width=\textwidth]{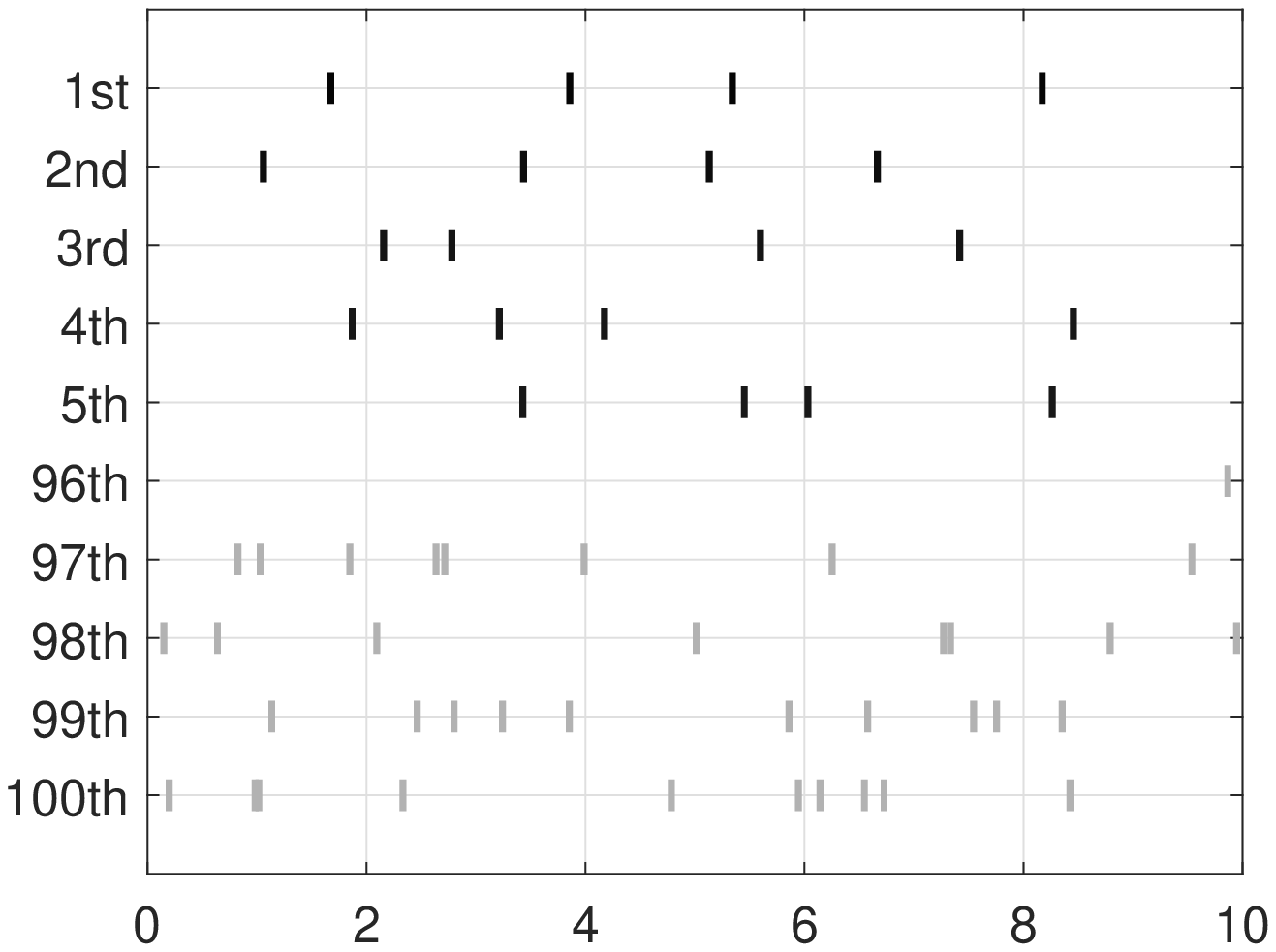}
    \caption{$r$=10}
    \label{HPPr10}
  \end{subfigure}%

\caption[]{Ranking HPP realizations. (a) 100 HPP realizations on [0, 10] with rate 0.4, where each row is one realization. (b) The  top $5$  and bottom $5$  ranked realizations using the conditional Dirichlet depth and $r=1$. (c) Same as (b) except that $r=10$. }
\label{fig3}
\end{figure}

Next, we will apply Dirichlet depth for HPP (Equation \eqref{eq:DirDepth HPP population}) as the conditional depth of the proposed depth framework (Equation \eqref{eq:depth}) on $100$ HPP realizations in interval $[0,10]$. The detailed procedure is:
(1)	Randomly generate $100$ HPP realizations in interval $[0, 10]$ with intensity rate $\lambda=0.4$ (the expected number of events should be $4$). 
(2)	Fit cardinality $|s|$ of sampled realizations into a Poisson model base on maximum Likelihood estimate of $\lambda$. Then use this model to compute probabilities $P(|s|=k)$ and normalize it as the first probability term $w(|s|=k)$ in Equation \eqref{eq:depth}. 
(3)	 Apply the Dirichlet depth for HPP as the conditional depth of Equation \eqref{eq:depth} to compute depth value of each realization.
The result is shown in Figure \ref{fig3}.

We can see that the depth values depend on both the probability term $w(|s|)$ and the conditional depth term, and the ranks vary with different choice of $r$. The deepest realization is the one with the largest weight ($|s|=4$) and evenly distributed (close to the center). When the value of $r$ changes  from 1 to 10, the first term $w(|s|)$ becomes more dominant, so that realizations with cardinality close to $4$ are more likely to be ranked on the top. Note that $r$ is a hyperparameter in the depth definition. In practice, one can adjust $r$ to balance the probability term and the conditional depth term for different purposes.  Alternatively, a cross-validation procedure may be applied to find an optimal value.


\subsection{Defining Dirichlet Depth for General Point Process} 
Based on Definition \ref{def:HPPdepth} for the HPP, we can now define the Dirichlet depth for general point process.  

\subsubsection{Definition} 

Defining Dirichlet depth for general point process is more challenging since there is no direct connection between Dirichlet distribution and point process (other than the HPP).  In this paper, we propose two different approaches for the definition:  (1) naturally extend Equation \eqref{eq:DirDepth HPP population} to general point process, and (2) transform the process to an HPP and then adopt Equation \eqref{eq:DirDepth HPP population}.  
At first, we extend the Dirichlet depth in Equation \eqref{eq:DirDepth HPP population} to a general point process in $[T_1,T_2]$ by defining the center of IETs as the conditional mean of the process.  The formal definition is given as follows. 

\begin{defn}\label{defn:DirDepth other population}
Given the cardinality $k$, assume the conditional mean of a point process in time $[T_1,T_2]$  as $\mu_{k}=(\mu_{1,k},\mu_{2,k},\cdots,\mu_{k,k})$. For an observed realization $s=(s_1, s_2,\cdots, s_k)$, set $s_0= \mu_{0,k}=T_1$ and $s_{k+1}= \mu_{k+1,k}=T_2$. If $\mu_{i,k} > \mu_{i-1,k}, i = 1, \cdots, k+1$, then the Dirichlet depth of $s$ is defined as:
\begin{equation} \label{eq:DirDepth other population}
D_c (s \mid |s|=k)=\prod^{k+1}_{i=1}(\frac{s_i-s_{i-1}}{\mu_{i,k} - \mu_{i-1,k}})^{\frac{\mu_{i,k} - \mu_{i-1,k}}{T_2-T_1}}.
\end{equation}
\end{defn}

For an HPP $s$ in $[T_1, T_2]$ conditioned on cardinality $|s|=k$, we have showed that its conditional mean is: $(\frac{T_2-T_1}{k+1}, \frac{2(T_2-T_1)}{k+1},\cdots,\frac{k(T_2-T_1)}{k+1})$. In this case, it is easy to verify that  Equation \eqref{eq:DirDepth other population} is simplified to 
Equation \eqref{eq:DirDepth HPP population}.  Therefore, Dirichlet depth for HPP is a special case of Equation \eqref{eq:DirDepth other population}. 
Note that the conditional depth value in Equation \eqref{eq:DirDepth other population} only depends on the conditional expectation of the process with the same cardinality, which can be estimated by the conditional sample mean given a collection of realizations. Hence, the sample version of Dirichlet depth can be obtained by replacing its conditional means with sample means. That is, we can write the {\bf sample Dirichlet depth} as:
\begin{equation}\label{eq:DirDepth other sample}
D_{c,n} (s \mid |s|=k)=\prod^{k+1}_{i=1}(\frac{s_i-s_{i-1}}{\overline{s}_{i,k}^{(n)}-\overline{s}_{i-1,k}^{(n)}})^{\frac{\overline{s}_{i,k}^{(n)}-\overline{s}_{i-1,k}^{(n)}}{T_2-T_1}},
\end{equation}
where $\overline{s}_{k}^{(n)}=(\overline{s}_{1,k}^{(n)},\overline{s}_{2,k}^{(n)},\cdots,\overline{s}_{k,k}^{(n)} )$ are the estimated sample means conditioned on the cardinality $k$ of $n$ observed realizations.

\subsubsection{Properties} 

For point processes in general, we also adopt the center as the conditional mean given cardinality $|s|=k$.  That is, $\theta_k= \mathbb{E}(s \mid |s| = k)$. 
We can show that the Dirichlet depth of Equation \eqref{eq:DirDepth other population} assigns the highest depth value $1$ to this center.  For HPP, this center is always $(\frac{T_2-T_1}{k+1},\frac{2(T_2-T_1)}{k+1},\cdots,\frac{k(T_2-T_1)}{k+1})$, regardless of the constant intensity rate $\lambda$.  However, in general, there is no closed-form expression of the conditional mean if the conditional intensity of the process is unknown. Indeed, we can prove that Properties P-1 to P-4 are satisfied by the general Dirichlet depth in Equation \eqref{eq:DirDepth other population}.  The detailed proof is omitted due to its similarity to the HPP case.

%
%
%

Now we examine the time invariance property.  Let $p = (p_{1}$, $\cdots, p_{k})$ be a random point process realization with $k$ events.  Then the conditional mean is $\mathbb E(p \mid |p|=k) = (\mathbb E(p_1 \mid |p|=k), \cdots, \mathbb E(p_k \mid |p|=k))$.  Under time warping $\gamma \in \Gamma$, a point process $s=(s_1, \cdots, s_k)$ will become $\gamma(s)=(\gamma(s_1)$, $\cdots, \gamma(s_k))$.  Similarly, the conditional means will become
$\mathbb E(\gamma(p) \mid |p|=k) = (\mathbb E(\gamma(p_1) \mid |p|=k), \cdots, \mathbb E(\gamma(p_k) \mid |p|=k)).$  To simplify the notation on the conditional means, we let $\mu_{i,k} = \mathbb E(p_i \mid |p|=k)$ and $\mu_{i,k,\gamma} = \mathbb E(\gamma(p_i) \mid |p|=k), i = 1, \cdots, k$ and $\mu_{0,k} = \mu_{0,k,\gamma} = T_1, \mu_{k+1,k} = \mu_{k+1,k,\gamma} = T_2.$ 
 If we include the conditional means in the definition, the Dirichlet depth on the transformed point process is:
\begin{align*}
& D_c (\gamma(s) \mid |\gamma(s)|=k, \{\mu_{\cdot, k,\gamma}\})=\prod^{k+1}_{i=1} \left (\frac{\gamma(s_i)-\gamma(s_{i-1})}{\mu_{i,k,\gamma} - \mu_{i-1,k,\gamma}} \right ) ^{\frac{\mu_{i,k,\gamma} - \mu_{i-1,k,\gamma}}{\gamma(T_2)-\gamma(T_1)}} \\
& \neq  \prod^{k+1}_{i=1}(\frac{s_i-s_{i-1}}{\mu_{i,k} - \mu_{i-1,k}})^{\frac{\mu_{i,k} - \mu_{i-1,k}}{T_2-T_1}} = 
D_c (s \mid |s|=k, \{\mu_{\cdot, k}\}). 
\end{align*}
The inequality holds because the time warping in general is a nonlinear transformation. That is, Property P-5 does not hold for the Dirichlet depth in Definition \ref{defn:DirDepth other population}.

\subsubsection{Bootstrapping estimation} 

Dirichlet depth in Equation \eqref{eq:DirDepth other population} relies heavily on the conditional means. For point process in general, there are no closed-forms for the conditional means if conditional intensity function is unknown.  In practice, given a set of point process realizations, we can apply the sample version Equation \eqref{eq:DirDepth other sample} to estimate the Dirichlet depth. However, for a given training data set, the sample size usually is not sufficiently large to result in a proper estimation of the conditional mean for each cardinality $|s|$. Here we propose a bootstrapping approach to address this issue.

Given a data set of point process realizations $p_1, p_2, \cdots, p_n$, where $p_i$ is a vector in $\mathbb{R}^{|p_i|}$ for $i=1, 2, \cdots, n$. In general, those vectors do not have the same dimension, and therefore it is not possible to take an average to compute the conditional sample mean as we need in sample Dirichlet depth.  To address this issue, we propose a bootstrap method to resample each realization $p_i$ such that the resampled realizations $p'_i$ has the desired dimension $k$. Then we can effectively estimate the conditional sample mean given cardinality $|s|=k$ by simply taking an average. The detailed steps are listed in Algorithm~\ref{alg: bootstrapping} as follows.

\begin{algorithm} 
\caption{Bootstrapping method to estimate conditional means}
\begin{algorithmic} 
\REQUIRE Given a sequence of realizations of point process $p_1, p_2, \cdots, p_n$
\STATE Combine all events of $p_1, p_2,\cdots, p_n$ together, $p_{com}=(p_1,p_2, \cdots, p_n)$
\FOR{ $k=1 \ to \ max(|p_i|) $}
\FOR{$ i=1 \ to \ n$ }
\STATE if $|p_i| \ge k$, then uniformly randomly delete $|p_i|-k$ events in $p_i$.
\STATE Otherwise, add $k-|p_i|$ by uniformly re-sampling from $p_{com}$ with replacement.
\ENDFOR
\STATE Denote $n$ resampled realizations as $p'_{1,k},p'_{2,k},\cdots, p'_{n,k}$, and then the estimated conditional mean is:
\STATE $$\overline{s}_{k}^{(n)} =\frac{1}{n}\sum_{i=1}^{n}p'_{i,k}$$
\ENDFOR
\RETURN $[\overline{s}_{k}^{(n)}]_{k=1}^{max(|p_i|)}$
\end{algorithmic}
\label{alg: bootstrapping}
\end{algorithm}


\subsection{Alternative Definition of Dirichlet Depth for General Point Process} 
The time warping transformation allows all events in a point process freely move in the given domain, while remaining the order of them. 
Ideally, the center-outward ranks of a set of point processes will remain the same if the same transformation is applied on all of them.  The invariance under such transformation is of great interest.  However, we have shown that the Dirichlet depth in Equation \ref{eq:DirDepth other population} does not have such invariance.  
In this subsection, we seek for an alternative definition of the Dirichlet depth to satisfy this property. 

\subsubsection{Definition} 

 Note that we have defined the depth for HPP.  For any point process, if we can find a way to transform it to an HPP, then the notion of Dirichlet depth can be directly applied.  Actually, such transformation can be done using the well-known {\bf time re-scaling theorem} \citep{Brown01}.  
 Basically, the theorem states that any point process with an integrable conditional intensity function can be converted into an HPP \citep{Papangelou72, Karr91}: Let $T_1< s_1 < s_2< \cdots < s_k \leq T_2$ be a realization from a point process with a conditional intensity function $\lambda(t|H_t) > 0$ for all $t \in (T_1,T_2].$ Then, the sequence $ \Lambda(s_i) = \int_{T_1}^{s_i} \lambda(t|H_t) dt, i = 1, \cdots, k$ is a Poisson Process with the unit rate in $(0, \Lambda(T_2)]$.

By applying this theorem, the notion of Dirichlet depth can be extended to general point processes. For a point process with known conditional intensity function, we can apply the time re-scaling theorem to convert it into an HPP in $[0,1]$, and then use Equation \eqref{eq:DirDepth HPP population} to compute its Dirichlet depth. Here we propose an alternative definition of the Dirichlet depth, referred to as time-rescaling-based (or TS-based) Dirichlet depth, as follows:

\begin{defn} \label{defn:TS_based DirDepth}
For a point process in time $[T_1,T_2]$ with conditional intensity function $\lambda(t \mid H_t)>0 $ and $\Lambda(t)=\int_{T_1}^t \lambda(u \mid H_u) du$, we define a time-rescaling-based conditional Dirichlet depth of a realization $s=(s_1,s_2,....,s_k)$ as: 
\begin{equation} \label{eq:TS-based DirDepth}
D_{c-TS} (s \mid |s|=k)
=(k+1)\prod^{k+1}_{i=1}(\frac{\Lambda(s_i)-\Lambda(s_{i-1})}{\Lambda(T_2)})^{\frac{1}{k+1}}, 
\end{equation}
where $s_0=T_1 \ and \ s_{k+1}=T_2$, 
\end{defn}

We can verify that the sequence $(\frac{\Lambda(s_1)}{\Lambda(T_2)}, \cdots, \frac{\Lambda(s_k)}{\Lambda(T_2)})$ follows an HPP in $[0,1]$ with intensity $\Lambda(T_2)$. For point processes without history dependence such as an inhomogeneous Poisson process (IPP), the re-scaled HPP realization will be distributed in a fix time interval $[0,\Lambda(T_2)]$. However, for point processes with history dependence, the re-scaled HPP realizations have different time length. A normalization to the interval $[0,1]$ by dividing $\Lambda(T_2)$ will help make comparison across realizations. 
Note that Definition \ref{defn:TS_based DirDepth} is not IET-based with respect to the original point process, and therefore there is no notion of simplex. This is fundamentally different from Definitions \ref{defn:DirDepth HPP population} and \ref{defn:DirDepth other population}.
Moreover, for an HPP in $[T_1, T_2]$ with constant rate $\lambda$, $\Lambda(t) = \lambda(t-T_1)$.   Then Equation  \eqref{eq:TS-based DirDepth} is simplified to 
$$(k+1)\prod^{k+1}_{i=1}(\frac{\lambda(s_i-s_{i-1})}{\lambda(T_2-T_1)})^{\frac{1}{k+1}} = 
(k+1)\prod^{k+1}_{i=1}(\frac{s_i-s_{i-1}}{T_2-T_1})^{\frac{1}{k+1}}.
$$
Therefore, the TS-based Dirichlet depth also generalizes Definition  \ref{def:HPPdepth} for HPP.  

\subsubsection{Properties}

We examine the 5 properties of the time re-scaling based Dirichlet depth $D_{c-TS}$ (Equation \ref{eq:TS-based DirDepth}).  We will show that P-1, P-4, and P-5 still hold.  

{\bf P-1: Continuity and vanishing at the boundary:}
 The detailed proof of P-1 is shown in Part C of the Supplementary Materials. 

{\bf P-2: Maximality at the center:}  The function $\Lambda(\cdot)$ varies with respect to each point process. Hence, in general, there could be multiple maxima in the TS-based Dirichlet depth and a unique center will not exist.  However, in the special case of  IPP, $\Lambda$ is a deterministic function for all processes.  Let $\theta_k=(\frac{\Lambda(T_2)}{k+1},\frac{2\Lambda(T_2)}{k+1},\cdots,\frac{k\Lambda(T_2)}{k+1})$ denote the center of an HPP on $[0, \Lambda(T_2)]$ with cardinality $k$.  It is easy to see that $ D_{c-TS}(\Lambda^{-1}(\theta_k) \mid |\theta_k|=k)= \sup_{s \in S_k} D_{c-TS}(s \mid |s|=k ) = 1$, and this center 
$\Lambda^{-1}(\theta_k)$ is unique.

{\bf P-3: Monotonicity relative to the deepest point:}  As the center may not be unique, the monotonicity cannot hold. 


{\bf P-4: Scale and shift invariance:}  The proof is similar to the one in the HPP case. 

{\bf P-5, Time warping invariance:}  Under time warping $\gamma \in \Gamma$, a point process $s=(s_1, s_2, \cdots, s_k)$ will become $\gamma(s)=(\gamma(s_1),\gamma(s_2)$, $\cdots, \gamma(s_k))$.  Similarly, we find that the transformed cumulative conditional intensity function 
$\Lambda_\gamma = \Lambda \circ \gamma^{-1}$. 
If we include the cumulative conditional intensity function in the definition of Dirichlet depth,  we have 
\begin{align*}
&D_{c-TS}(\gamma(s) \mid |s|=k, \Lambda_\gamma) =  
(k+1)\prod^{k+1}_{i=1}(\frac{\Lambda_\gamma(\gamma(s_i))-\Lambda_\gamma(\gamma(s_{i-1}))}{\Lambda_\gamma(\gamma(T_2))})^{\frac{1}{k+1}}, 
\\
&=(k+1)\prod^{k+1}_{i=1}(\frac{\Lambda(s_i)-\Lambda(s_{i-1})}{\Lambda(T_2)})^{\frac{1}{k+1}}
= D_{c-TS}(s \mid |s|=k, \Lambda).
\end{align*}
The detailed proof is given in Part D of the Supplementary Materials.

Comparing two different methods (Definitions \ref{defn:TS_based DirDepth} and \ref{defn:DirDepth other population}) of defining Dirichlet depth for general point process, Definition \ref{defn:DirDepth other population} seems easier to apply in practice. If the conditional intensity function is known, the proposed TS-based definition is expected to have an effective center-outward ranking.  However, the conditional intensity function is often unknown, particularly in practical use.  In fact, perhaps the most challenging part of the TS-based Dirichlet depth in Equation \eqref{eq:TS-based DirDepth} is to properly estimate the conditional intensity. 

As a summary, we list the properties in the three Dirichlet depths (in Definitions  \ref{defn:DirDepth HPP population}, \ref{defn:DirDepth other population}, and \ref{defn:TS_based DirDepth})
in  Table \ref{tab:sum}, where ``T'' denotes ``true'' and ``F'' denotes ``false''.  

\begin{table}[h]
\caption{Properties of the Proposed Dirichlet depths}
\centering
\begin{tabular}{c|c|c|c|c|c}
\hline\hline 
Dirichlet Depth Method & P-1 & P-2 &P-3 & P-4 & P-5 \\ [0.5ex] 
\hline\hline 
 Depth on HPP  & T & T & T & T & N/A \\
\hline 
 Depth on Point Process & T & T & T & T & F \\
\hline
TS-based Depth on Point Process & T & F \ (T for IPP) & F & T & T \\
\hline 
\end{tabular}
\label{tab:sum}
\end{table}

\subsubsection{Illustration} 

\begin{figure}[!h]
  \centering
  \begin{subfigure}[b]{0.33\textwidth}
    \includegraphics[width=\textwidth]{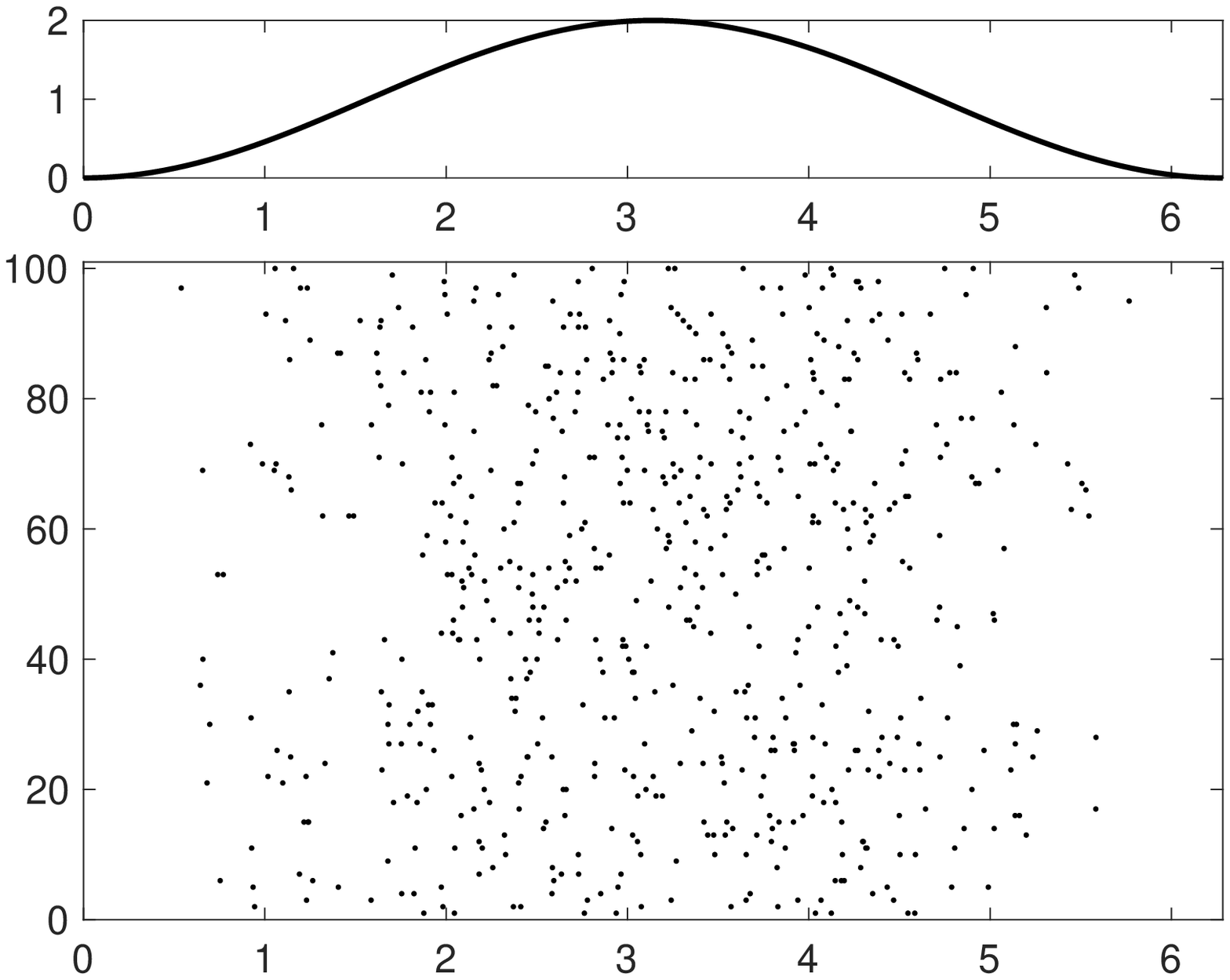}
    \caption{IPP}
    \label{IPP_ori}
  \end{subfigure}%
    \begin{subfigure}[b]{0.33\textwidth}
    \includegraphics[width=\textwidth]{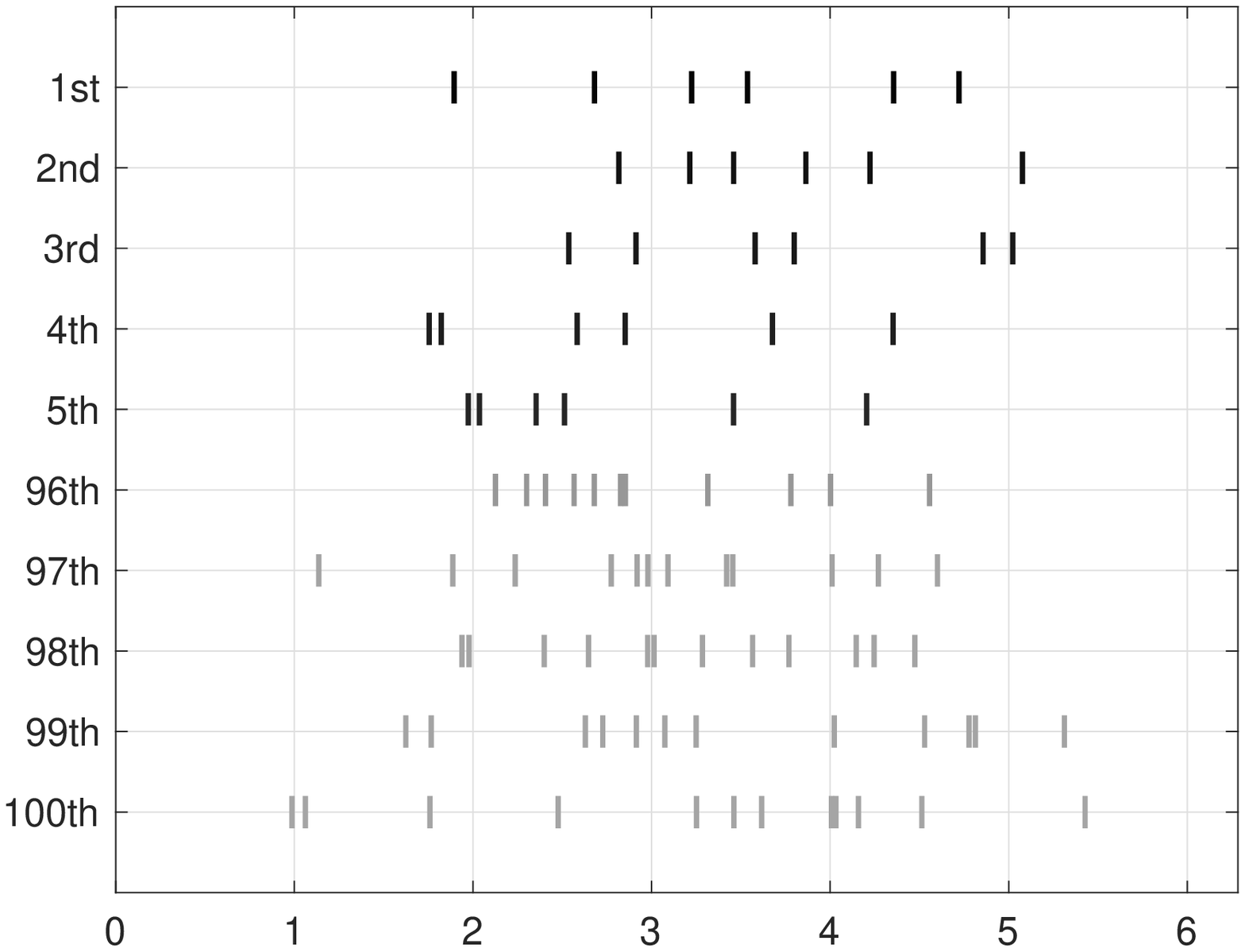}
    \caption{Dirichlet, r=1}
    \label{IPP_M2r1}
    \end{subfigure}%
     \begin{subfigure}[b]{0.33\textwidth}
    \includegraphics[width=\textwidth]{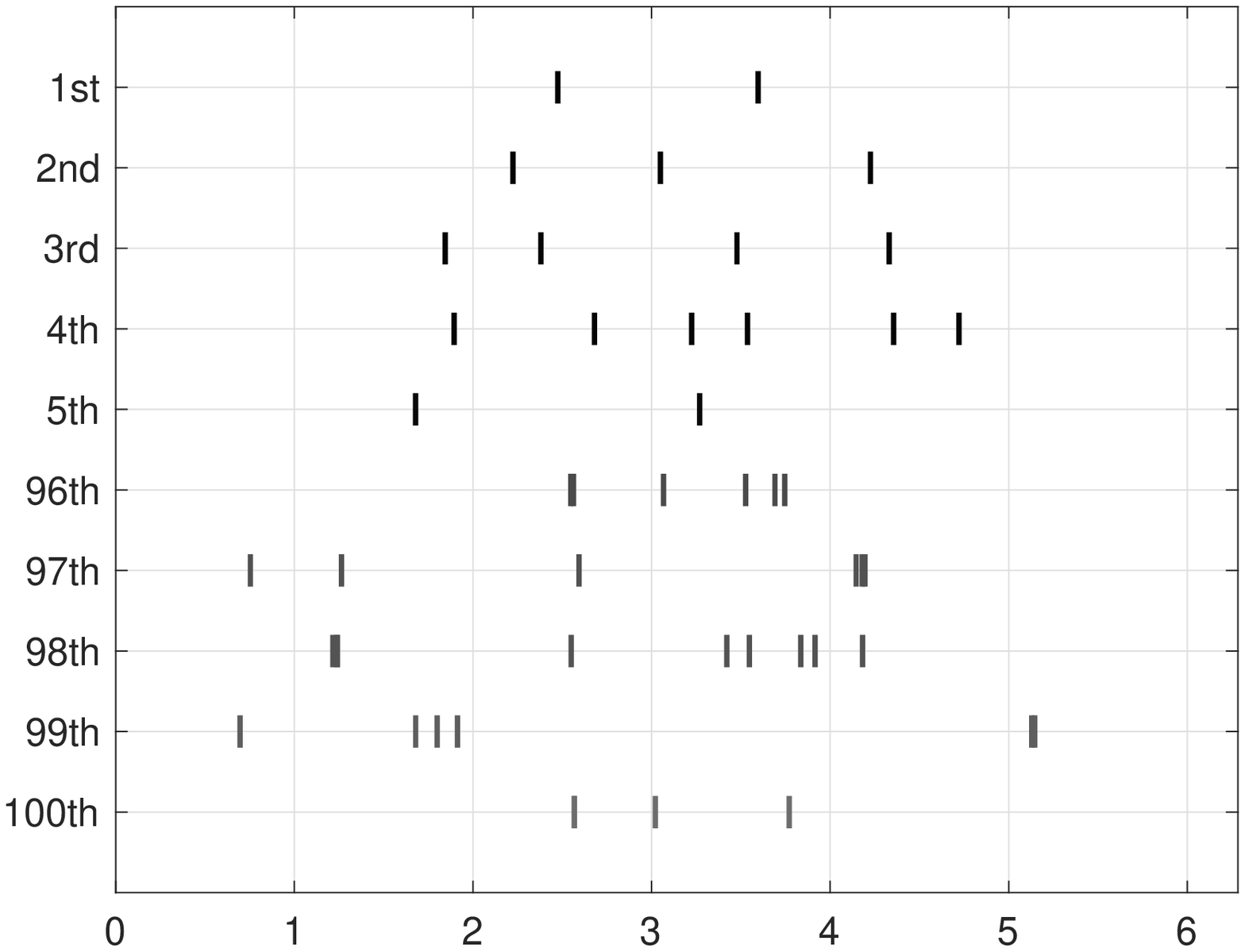}
    \caption{Dirichlet, r=0.01}
    \label{IPP_M2r001}
    \end{subfigure}%
\\
    \begin{subfigure}[b]{0.33\textwidth}
    \includegraphics[width=\textwidth]{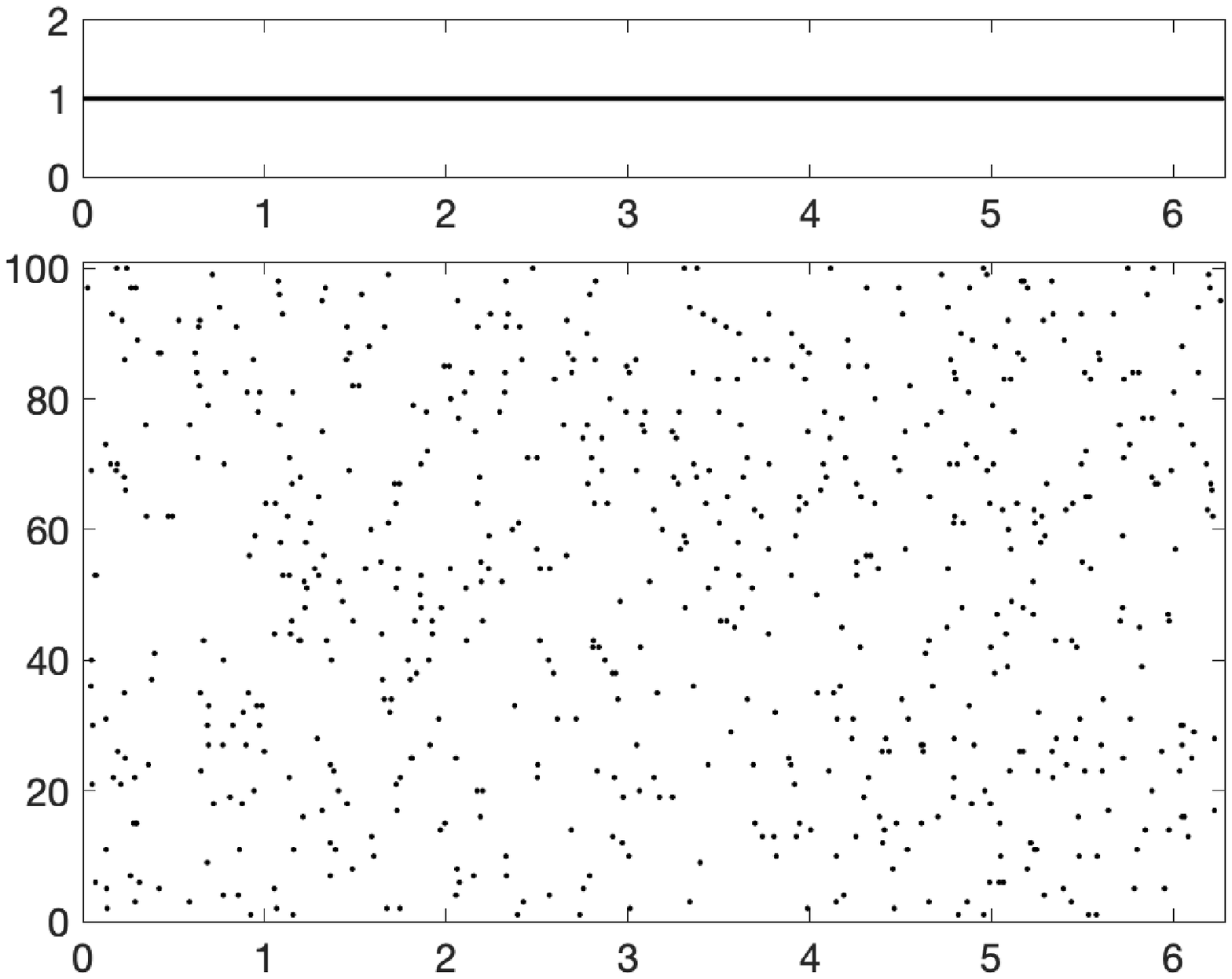}
    \caption{Re-scaled}
    \label{IPP_TS}
  \end{subfigure}%
  \begin{subfigure}[b]{0.33\textwidth}
    \includegraphics[width=\textwidth]{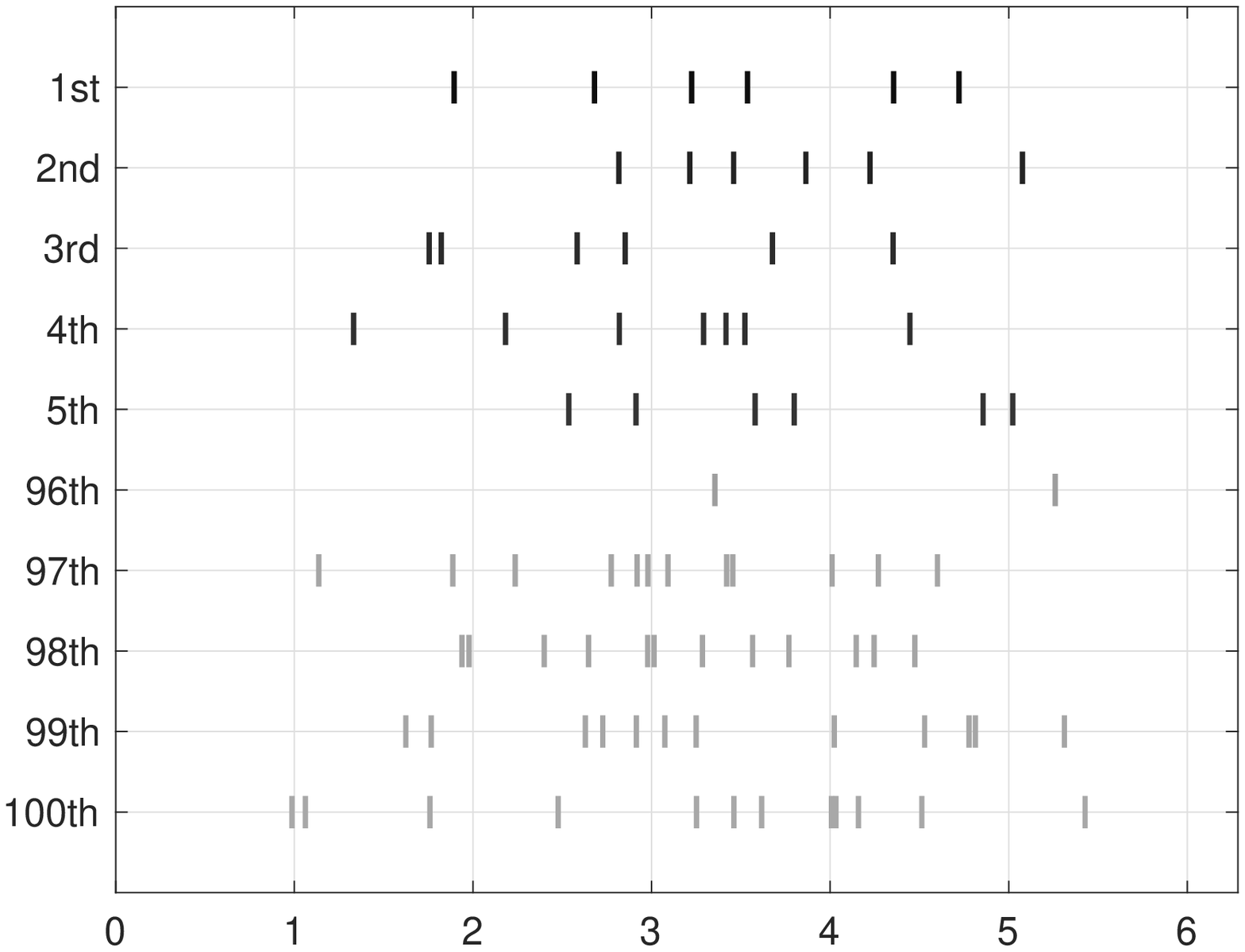}
    \caption{TS-based, r=1}
    \label{IPP_TSDDr1}
  \end{subfigure}%
    \begin{subfigure}[b]{0.33\textwidth}
    \includegraphics[width=\textwidth]{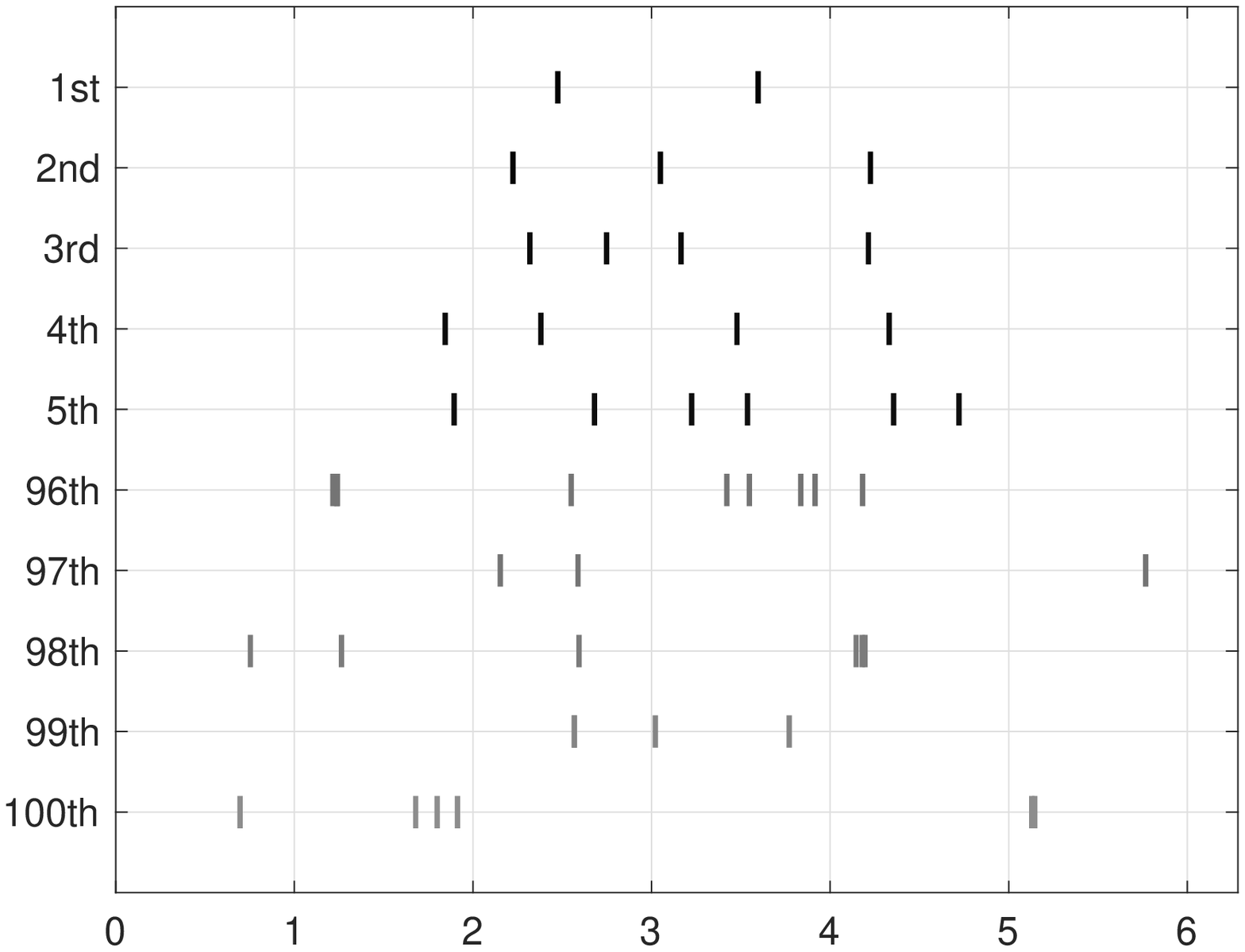}
    \caption{TS-based, r=0.01}
    \label{IPP_TSDDr001}
  \end{subfigure}%
  \caption[]{Dirichlet depth for an IPP. (a) 100 IPP realizations (bottom) on $[0,2 \pi]$ with the intensity function $\lambda(t)=1-cos(t)$ (top). (b) 
 The top 5 and bottom 5 ranked realizations with the sample Dirichlet depth and $r = 1$. (c) Same as (b) except for $r=0.01$.  (d) TS-based transferred realizations (bottom) and its intensity function (top). (e) Same as (b) except for  the TS-based Dirichlet depth.  (f) Same as (e) except for $r=0.01$.}
\label{figIPP}
\end{figure}

In this subsection, we demonstrate the proposed Dirichlet depths using a simulated inhomogeneous Poisson process (IPP). We randomly generate $100$ IPP realizations on $[0,2\pi]$ with intensity function $\lambda(t)=1-cos(t)$. The generated realizations are shown in Figure \ref{figIPP}(a). The total intensity is $\Lambda=\int_0^{2 \pi}\lambda(t)dt = 2\pi$, and therefore the probability of cardinality $P(|s|)$ reaches its maximum at $|s|=6$.


Given a sample set of a point process, we need to estimate the conditional means in order to apply sample Dirichlet depth and need to estimate the intensity function for TS-based Dirichlet depth. In this example, we use Algorithm \ref{alg: bootstrapping} to estimate the conditional means.  The intensity function of an IPP can be easily estimated with training samples. 

We will also need to estimate the probability $P(|s|)$ in order to compute the first probability term $w(|s|)$ in Equation \eqref{eq:depth}. In this illustration, $P(|s|)$ is estimated by the MLE algorithm based on Poisson distribution and remains the same under different depth functions. Moreover, the weight parameter $r$ is introduced to balance the importance of the first probability term and conditional depth. For illustrative purposes, we set $r$ to two different values of 1 and 0.01.  The ranking result is shown in Figure \ref{figIPP}.
 
Comparing Panels (b) and (e) (where $r=1$), we can see that the ranking results based on sample Dirichlet depth function and TS-based Dirichlet depth function are very similar -- four out of five top-ranked realizations are the same. Similarly, four out of five bottom-ranked realizations are the same as well. We also compare Panels (c) and (f) where $r=0.01$.  Although the overall ranks changed dramatically from where $r=1$, both methods agree on the four out of five deepest realizations, and four out of five shallowest realizations.

\begin{figure}[ht]
  \centering
  \begin{subfigure}[b]{0.35\textwidth}
    \includegraphics[width=\textwidth]{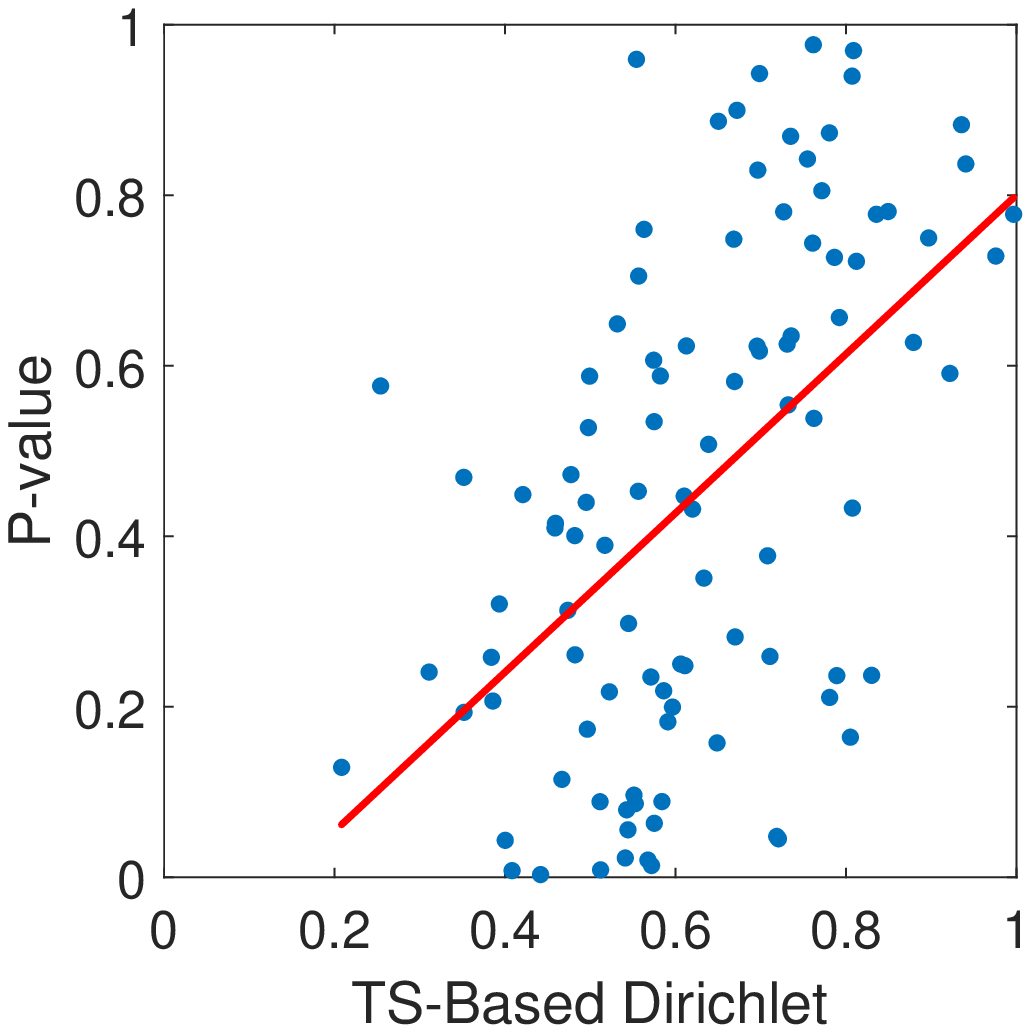}
    \caption{TS-based Dirichlet depth}
    \label{ks_ts}
  \end{subfigure}%
    \begin{subfigure}[b]{0.35\textwidth}
    \includegraphics[width=\textwidth]{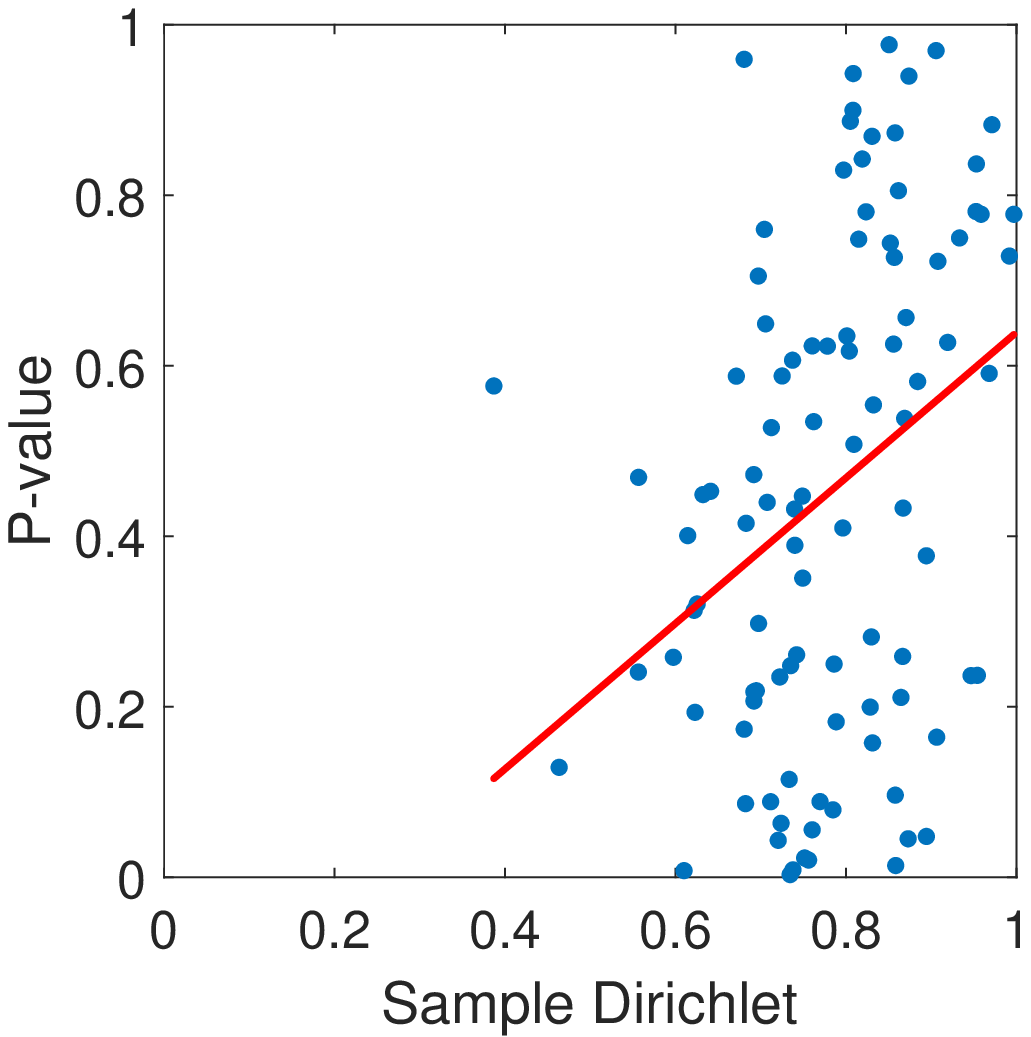}
    \caption{Sample Dirichlet depth}
    \label{ks_sample}
    \end{subfigure}%
  \caption[]{``Goodness-of-fit'' of TS-based Dirichlet depth and sample Dirichlet depth. (a) Plot of TS-based Dirichlet depth values vs. P-values. The solid line indicates the fitted linear regression. (b) Same as (a) except for sample Dirichlet depth values vs. P-values. }
\label{fig5}
\end{figure}

We have seen that both Dirichlet depths can identify the most "typical" realizations whose distributions closely resemble the true intensity function. In order to illustrate the relationship between the Dirichlet depth and the goodness-of-fit, we treat each realization as a set of i.i.d. sampling points from the density function $f(t)=\frac{\lambda(t)}{\Lambda(2\pi)}$, and plot its P-values from the Kolmogorov-Smirnov (KS) test vs. its conditional Dirichlet depths in Figure \ref{fig5}.


We can see that both sample Dirichlet depth and TS-based Dirichlet depth are positively associated with the P-value of the KS test.  This indicates more typical realizations correspond to larger Dirichlet depth values.  Therefore, the proposed Dirichlet depths provide an alternative measurement for the ``goodness-of-fit'' of the given observations on the intensity function.



\section{Asymptotic Theory}
\label{sec:asymptotic}

In this section, we will investigate the asymptotic behavior of the sample depth function for point process based on our proposed framework (Equation \eqref{eq:depth}). In practice, given a set of realizations, we need to estimate the probability term $w(|s|)$ and Dirichlet depth $D_c(s\mid |s|)$.  $r$ is a pre-set hyperparameter.  
Given a sample set $S^{(n)}$ that contains $n$ realizations from a point process on interval $[T_1, T_2]$, the probability term $w(|s|=k)$ can be estimated by:
$$w_n(|s|=k) = \frac{\# \mbox{ processes in } S^{(n)} \mbox{ with } k \mbox{ events}}
{\max_{0\le k \le K} \{\# \mbox{ processes in } S^{(n)} \mbox{ with } k \mbox{ events}\}},$$
with a pre-determined $K \in \mathbb N$.  Basically, $w_n(|s|=k)$ is the empirical probability mass function, normalized with maximum being 1.
The conditional Dirichlet depth proposed in Equation \eqref{eq:DirDepth other population} can be estimated by the sample Dirichlet depth in Equation \eqref{eq:DirDepth other sample}. Then we have a sample version of Equation \eqref{eq:depth} $D_n(s)$ in the following form: 
\begin{equation}
D_n(s)=w_n(|s|)^rD_{c,n} (s \mid |s|)
\label{eq:sampledepth}
\end{equation}

To simplify the theoretical derivation, we make the following three assumptions.  
\begin{enumerate}
\item The number of events in each process has a constant upper bound $K \in \mathbb N$, which can be arbitrarily large.
\item $w(|s|=k) > 0, k = 1, \cdots, K$.
\item $\mu_{i,k} > \mu_{i-1,k}, i = 1, \cdots, k+1, k = 1, \cdots, K$. 
\end{enumerate}
We have defined $ S_k= \{s=(s_1, \cdots, s_k) \in \mathbb R^k | T_1 \le s_1 \le \cdots \le s_k \le T_2\} $. Let $E^{(K)}=\bigcup^{K}_{k=0}S_k$. The depth function in Equation \eqref{eq:depth} on $E^{(k)}$ is a function $D:E^{(K)}\rightarrow[0,1]$.  
Our main asymptotic result is given as follows:
\begin{thm} \label{asymptotic}
For arbitrarily large $K \in \mathbb N$, let $s\in E^{(K)}$ be a point process realization in the time domain $[T_1,T_2]$.  If the three assumptions given above are satisfied, then
\begin{equation}
 \ \sup_{s\in E^{(K)}}|D_n(s)-D(s)|\rightarrow0 \ a.s. \ (as \ n\rightarrow \infty)
 \label{eq:th1}
 \end{equation}
Furthermore for $\alpha \in (0, 1]$, denote $D^{\alpha}\equiv\{s \in E^{(K)} \mid D(s)\geq\alpha\}$ and $D^{\alpha}_n \equiv \{s \in E^{(K)} \mid D_n(s)\geq\alpha\}$ as $\alpha-trimmed \ regions$.  Then for any $\epsilon \in (0, \min\{\alpha, 1-\alpha\})$, 
\begin{enumerate}
\item $D_n^{\alpha+\epsilon} \subset D^\alpha \subset D_n^{\alpha-\epsilon} $ for $n$ sufficiently large. 
\item $D_n^{\alpha} \rightarrow D^{\alpha}$ \ a.s.\ as \ $n \rightarrow \infty$ if $P(\{s \in E^{(K)}  \mid D(s)=\alpha\})=0$. 
\end{enumerate}
\end{thm}
The proof of Theorem \ref{asymptotic} is given in Part E of the Supplementary Materials. 




\section{Real Experimental Data}
\label{sec:real}
In this section, we will apply the proposed depths to study neural decoding problems in two spike train datasets, where spike trains can be naturally treated as point processes.  

\subsection{Motor Cortical Spike Trains}
\begin{figure}[thp]
  \centering
    \includegraphics[width=0.6\textwidth]{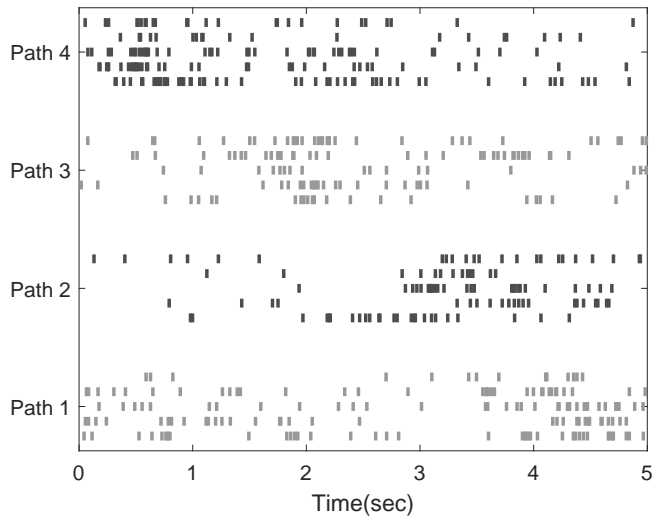}
  \caption[]{5 sample neural spike trains for each of 4 paths}
    \label{reals}
\end{figure}

We will at first perform a classification analysis on a set of motor cortical spike trains that was previously used in \cite{Wu13} and \cite{Liu2017}. In this experiment, researchers implanted a microelectrode array in the arm area of the primary motor cortex of a juvenile male macaque monkey to record neural spiking signals. The experiment subject was trained to perform a closed Square-Path task by moving a cursor to targets via contralateral arm movements in the horizontal plane. Each sequence of 5 targets defined a path, and there were four different paths in the task (depending on starting point).  Neural spike trains from single units were recorded during the behaviors. The dataset consists of 240 spike trains with 60 trains for each path, and the recording time was normalized to 5 seconds. Figure \ref{reals} shows 5 example spike trains for each path.  We take 30 trains in each path as the training data to estimate parameters in the depth function, and then use the other 30 trains as the test data to evaluate the depth values.  

To estimate the probability term $w(\cdot)$ in the proposed depth function, we assume that the cardinality $|s|$ follows a mixture of Poisson distribution.  The parameters in the model can be estimated via a classical EM algorithm, where the number of components in this mixture is determined via standard model selection methods.  

For classification analysis, one can compute the Dirichlet depth directly by using Equation \eqref{eq:DirDepth other sample} and the TS-based Dirichlet depth by using Equation \eqref{eq:TS-based DirDepth}. Computation of the sample Dirichlet depth requires estimation of conditional means.  Since the sample size is relatively small, we will use Algorithm \ref{alg: bootstrapping} to estimate the conditional means.  Also, TS-based Dirichlet depth requires estimation of intensity functions for each path.  Here we assume that there is no history dependence and use the IPP model to estimate the intensity function for each path.

\begin{figure}[th]
  \centering
    \includegraphics[width=0.7\textwidth]{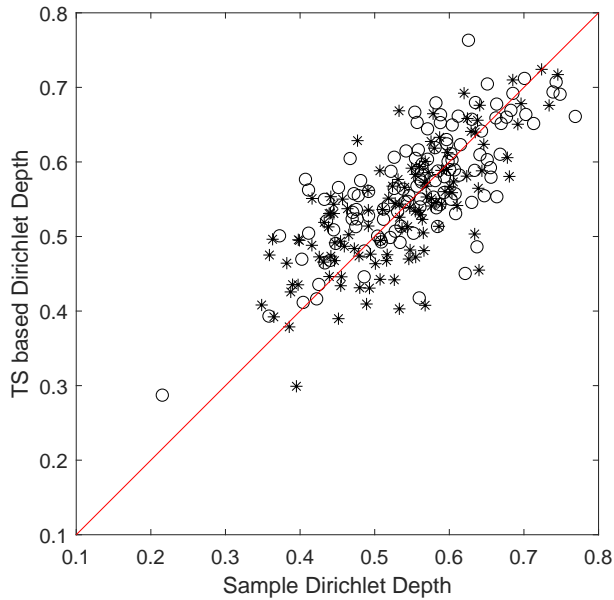}
  \caption[]{Plot of sample Dirichlet depth vs. TS based Dirichlet depth on both training set (asterisks) and test set (cicles).}
    \label{real_TSvsS}
\end{figure}

Ideally, the conditional depth generated by sample Dirichlet depth and TS-based Dirichlet depth should be consistent. From the plot of Dirichlet depths vs. TS-based Dirichlet depths (Figure \ref{real_TSvsS}), we see that the points are evenly spread around the red diagonal baseline, indicating that the two definitions produce similar ranking systems. 

Once the first probability term and the conditional depth function for each path are obtained from the training set, we can then estimate the depth values of spike trains of the test set for each path based on framework (Equation \eqref{eq:depth}).  We can then classify a test spike train to one of the four paths in which the depth value is the highest. 

To compare the classification performance with previous methods, we also apply two more approaches here: 1) generalized Mahalanobis depth method in \cite{Liu2017} and 2) likelihood method. We have discussed the Gaussian-kernel-based generalized Mahalanobis depth. The likelihood method, in this case, is based on the assumption that spike trains follow a Poisson process. We can estimate the likelihood of each testing spike train and classify it to the model in which the likelihood is the largest. The classification accuracy rates of all approaches are summarized in Table \ref{table:R1}. 

\begin{table}[h]
\caption{Comparison of classification performance}
\centering
\begin{tabular}{c c}
\hline\hline 
Method & Classification accuracy \\ [0.5ex] 
\hline 
\bf{TS-based Dirichlet depth }&\bf{0.90} \\ 
Generalized Mahalanobis depth & 0.87 \\
\bf{Sample Dirichlet depth} & \bf{0.74}\\
Likelihood method & 0.73\\ [1ex]
\hline 
\end{tabular}
\label{table:R1} 
\end{table}

The classification rate of depth function with TS-based Dirichlet depth is 0.90, which is slightly higher than the rate of 0.87 of the generalized Mahalanobis depth method.  The sample Dirichlet depth only has 0.74 classification rate, which is about the same level of accuracy as the likelihood method. We point out that the classification rate of sample Dirichlet depth depends on the bootstrapped conditional means; a better re-sampling method could achieve a better performance. In addition, the choice of intensity functions can affect the classification result as well.  Here we only assume that spike trains follows an inhomogeneous Poisson process. A more sophisticated framework of estimating intensity functions would affect the performance of classification. 

\subsection{Geniculate Ganglion Spike Trains}
We will use another spike train dataset to demonstrate the classification performance of the proposed framework. This dataset was previously used by \cite{Lawhern11} and \cite{Liu2017}, which contains spike trains of $6$ different clusters. In the experiment, adult male Sprague-Dawley rat’s geniculate ganglion tongue neurons were stimulated by $6$ different solutions: KCI (salty), CA (sour), NaCl (salty), QHCI (bitter), MSG (umami) and Sucr (sweet) for $10$ times each. The experiment consists of three time periods: 2-second pre-stimulus period, 2.5-second stimulus application period and 3-second post-stimulus period.

\begin{figure}[h]
  \centering
  \begin{subfigure}[b]{0.4\textwidth}
    \includegraphics[width=\textwidth]{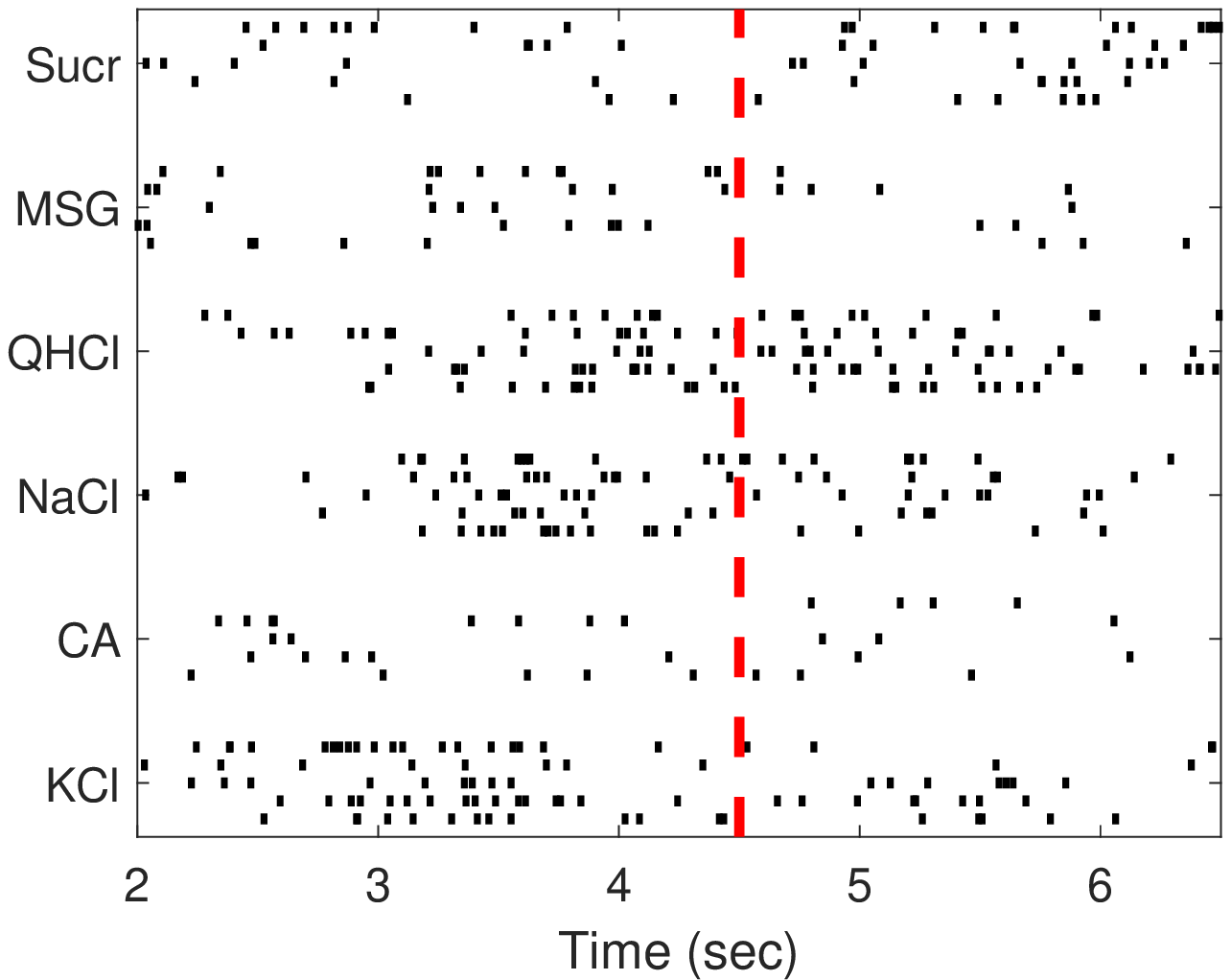}
    \caption{electrolyte generalist cell}
    \label{}
  \end{subfigure}%
    \begin{subfigure}[b]{0.4\textwidth}
    \includegraphics[width=\textwidth]{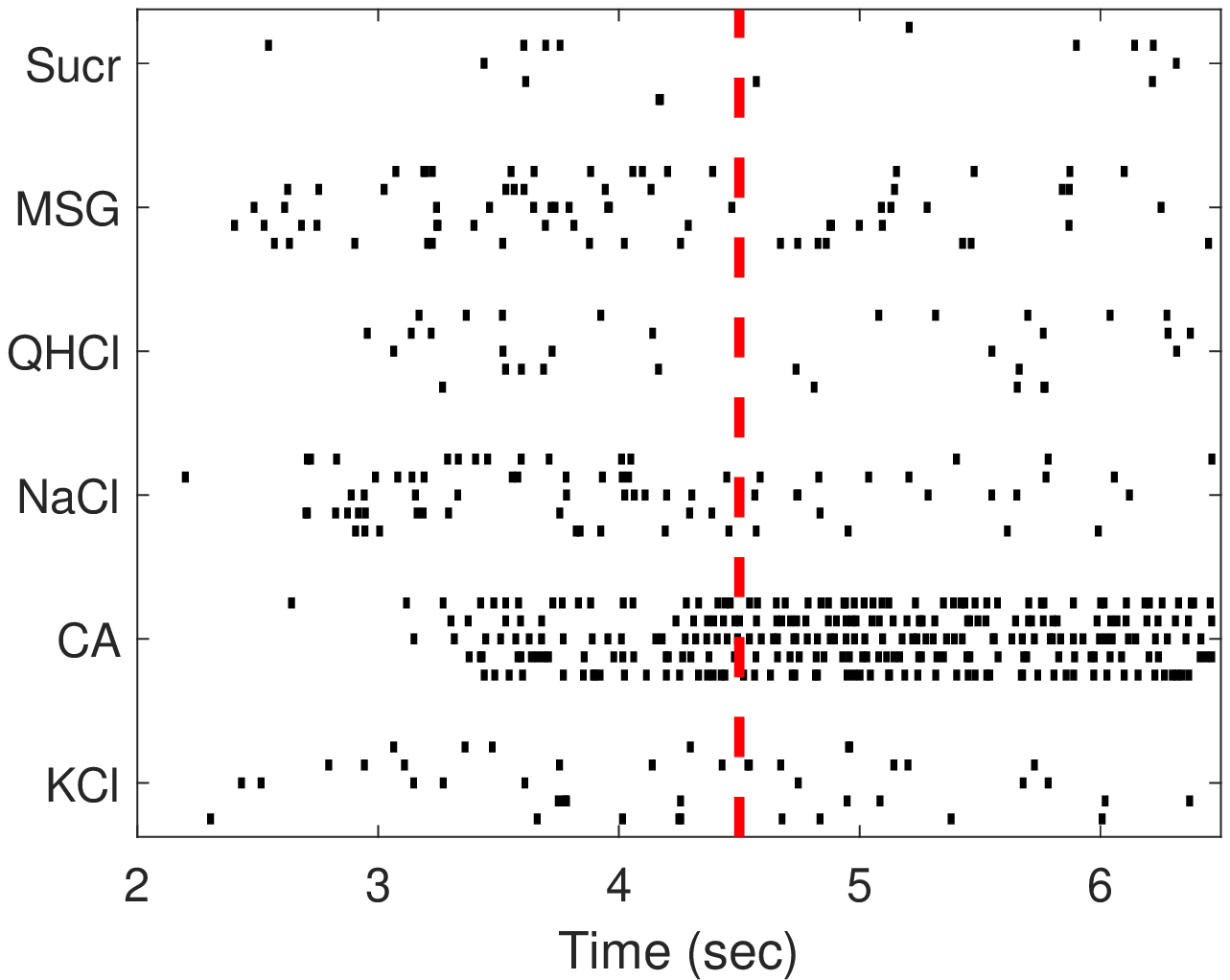}
    \caption{acid generalist cell}
    \label{}
    \end{subfigure}%
  \caption[]{Sample spike trains for different cells. (a) Sample spike trains of an electrolyte generalist cell.  Time interval before the vertical dashed line is the stimulus application period and that after is the post-stimulus period. (b) Same as (a) except for an acid generalist cell.}
  \label{tastedata}
\end{figure}

For illustrative purposes, we only use spike trains in the stimulus application period and the post-stimulus period, and only select two typical neurons cells: one electrolyte generalist cell and one acid generalist cell. For each cell, we take $5$ spike trains for each of $6$ different tastes to train, and another $5$ spike trains to perform classification task. That is, $60$ spike trains are been selected for each cell. The spike trains of the training set with respect to the $6$ different solutions from those two cells are shown in Figure \ref{tastedata}.

\begin{table}[h]
\caption{Comparison of classification performance}
\centering
\begin{tabular}{c c c}
\hline\hline 
Method & electrolyte generalist cell & acid generalist cell\\ [0.5ex] 
\hline 
\bf{Sample Dirichlet depth }&\bf{0.73} &\bf{0.83}\\ 
Generalized Mahalanobis depth & 0.70&0.76 \\
Likelihood method & 0.47&0.33\\ [1ex]
\hline 
\end{tabular}
\label{table:R2} 
\end{table}

We can see that the number of spikes is very different in (b), and not very different in (a). Hence we can predict that for classifying electrolyte generalist cell, the first probability term $w(|s|)$ will not be a very important factor, but can play an important role for classifying acid generalist cell. For electrolyte generalist cell, we set $r$ to be a small value of $0.3$, and for acid generalist cell $r$ is set as a larger value of $1.2$. In this example, only sample Dirichlet depth is used as the conditional depth function in Equation \eqref{eq:depth}.  The TS-based Dirichlet depth is omitted since the sample size is too small to have a robust estimate of the intensity function. Table  \ref{table:R2}  shows the result of the classification rate.  We can see from the classification result that our proposed sample Dirichlet depth is the best among three methods. In particular, given the extremely small sample size (5 in each taste), the classification performance is indeed robust and accurate. 

\section{Summary}
\label{sec:summary}
In this paper, we have proposed a new framework to measure depth for point process observations.  The proposed depth includes three components: 1) normalized probability of the number of events, 2) conditional depth given the number of events, and 3) the weight parameter.  Our study emphasizes on the definitions of the new conditional depth, namely the Dirichlet depth, for point process and important mathematical properties. The depth is at first defined for the classical homogeneous Poisson process by using the equivalent inter-event time representation. For general point process, we propose two different definitions: one is a direct generalization on the homogeneous Poisson case, and the other is based on the well-known time re-scaling theorem.  We examine the mathematical properties for each of these depths and provide a theoretical investigation on the asymptotics on the sample Dirichlet depth. 
Moreover, we apply the proposed depth functions to neural decoding problems in two different datasets. The result indicates that the proposed framework provides a proper center-outward rank and the new method has superior decoding performance to the previous methods. 

The new Dirichlet depths are simply based on the basic Dirichlet distribution and there are a lot of potential improvements for future exploration.  For example, we may change the power $\frac{1}{k+1}$ to make the depth more concentrated or dispersed.  This procedure may only slightly change the ranks of each sample, but will have clear effect on classification performance if the depth values are used across multiple samples. 

We point out that the Dirichlet depth is a new approach to define the conditional depth for point process.  To the best of our knowledge, no other methods  have been proposed to study this problem. Therefore, more in-depth topics, such as the shape of depth contours and trimmed regions in a high dimension, can be further explored. For practical application, we have only investigated the classification performance by the proposed depth framework. Other applied topics, such as clustering and outliers detection, can also be studied in the future.  Finally, the weight parameter $r$ is pre-set in this paper.  We will explore classical statistical methods such as cross-validation or generalized cross-validation to search for optimal values in practical use.

\bibliographystyle{agsm}

\bibliography{reference}

@article{Barnett76,
 author = {V. Barnett},
 journal = {Journal of the Royal Statistical Society. Series A (General)},
 number = {3},
 pages = {318--355},
 publisher = {[Royal Statistical Society, Wiley]},
 title = {The ordering of multivariate data},
 volume = {139},
 year = {1976}
}

@Article{Oja83,
  author={Oja, Hannu},
  title={{Descriptive statistics for multivariate distributions}},
  journal={Statistics \& Probability Letters},
  year=1983,
  volume={1},
  number={6},
  pages={327-332},
  month={October},
}

@article{Fraiman99,
author = {Fraiman, Ricardo and Meloche, Jean and García-Escudero, Luis and Gordaliza, Alfonso and He, Xuming and Maronna, Ricardo and Yohai, Victor and Sheather, Simon and Mckean, Joseph and G. Small, Christopher and Wood, Andrew},
year = {1999},
month = {12},
pages = {255-317},
title = {Multivariate L-estimation},
volume = {8},
journal = {Test}
}

@article{Zuo2000,
author = "Zuo, Yijun and Serfling, Robert",
fjournal = "The Annals of Statistics",
journal = "Ann. Statist.",
month = "04",
number = "2",
pages = "461--482",
publisher = "The Institute of Mathematical Statistics",
title = "General notions of statistical depth function",
volume = "28",
year = "2000"
}

@article{Nieto2016,
author = "Nieto-Reyes, Alicia and Battey, Heather",
fjournal = "Statistical Science",
journal = "Statist. Sci.",
month = "02",
number = "1",
pages = "61--79",
publisher = "The Institute of Mathematical Statistics",
title = "A topologically valid definition of depth for functional data",
volume = "31",
year = "2016"
}

@article{Liu2017,
author = "Liu, Shuyi and Wu, Wei",
fjournal = "The Annals of Applied Statistics",
journal = "Ann. Appl. Stat.",
month = "06",
number = "2",
pages = "992--1010",
publisher = "The Institute of Mathematical Statistics",
title = "Generalized Mahalanobis depth in point process and its application in neural coding",
volume = "11",
year = "2017"
}

@Article{Liu90,
	author    = {R. Liu},
	title     = {On a notion of data depth based on random simplices},
	journal   = {The Annals of Statistics},
	year      = {1990},
	volume    = {18},
	number    = {1},
	pages     = {405-414}
}

@Article{Wu13,
	author    = {Wu, W. and Srivastava, A.},
	title     = {Estimating Summary Statistics in the Spike-train Space},
	journal   = {Journal of Computational Neuroscience},
	year      = {2013},
	volume    = {34},
	number    = {},
	pages     = {391-410}
}

@Article{Tukey75,
	author    = {Tukey, J. W.},
	title     = {Mathematics and picturing data},
	journal   = {International Congress of Mathematicians},
	year      = {1975},
	volume    = {2},
	number    = {},
	pages     = {523-531}
}

@Article{Mosler12,
	author    = {Mosler,K. and Polyakova, Y.},
	title     ={General Notions of Depth for Functional Data},
	journal   = {Manuscript},
	year      = {2012},
	volume    = {1981},
	number    = {},
	pages     = {}
}

@Article{Pintado09,
	author    = {S. Lopez-Pintado and J. Romo},
	title     ={On the Concept of Depth for Functional Data},
	journal   = {Journal of American Statistical Association},
	year      = {2009},
	volume    = {104},
	number    = {486},
	pages     = {718-734}
}

@Article{Liu93,
	author    = {Liu, R. Y. and Singh, K.},
	title     ={A quality Index Based on Data Depth and Multivariate Rank Tests},
	journal   = {Journal of the American Statistical Association},
	year      = {1993},
	volume    = {88},
	number    = {},
	pages     = {252-260}
}

@Article{Lawhern11,
	author    = {V. Lawhern and A. A, Nikonov and W. Wu and R. J. Contrares},
	title     = {Spike rate and spike timing contributions to coding taste quality information in rat periphery},
	journal   = {Frontiers in Integrative Neuroscience},
	year      = {2011},
	volume    = {5},
	number    = {},
	pages     = {1-14}
}

@Article{Brown01,
  author    =  {Brown, E. N. and Barbieri, R. and Ventura, V. and Kass, R. E. and Frank, L. M.},
  title     =  {The time-rescaling theorem and its application to neural spike train data analysis},
  journal   =  {Neural Computation},
  year =       {2001},
  volume    =  {14},
  pages     =  {325--346}
}

@article{Liu99,
   author = {Liu, Regina Y. and Parelius, Jesse M. and Singh, Kesar},
   fjournal = {The Annals of Statistics},
   journal = {Ann. Statist.},
   month = {06},
   number = {3},
   pages = {783--858},
   title = {Multivariate analysis by data depth: descriptive statistics, graphics and inference, (with discussion and a rejoinder by Liu and Singh)},
   volume = {27},
   year = {1999}
}

@article{Papangelou72,
   author = {Papangelou, F},
   year = {1972},
   month = {03},
   pages = {},
   title = {Integrability of Expected Increments of Point Processes and a Related Random Change of Scale},
   volume = {165},
   journal = {Transactions of the American Mathematical Society - TRANS AMER MATH SOC}
}

@book{Karr91,
  title={Point processes and their statistical inference, Second Edition,},
  author={Karr, A.},
  isbn={9780824785321},
  lccn={90028376},
  series={Probability: Pure and Applied},
  year={1991},
  publisher={Taylor \& Francis}
}

@article{zuo2000b,
author = {Zuo, Yijun and Serfling, Robert},
fjournal = {The Annals of Statistics},
journal = {Ann. Statist.},
month = {04},
number = {2},
pages = {483--499},
publisher = {The Institute of Mathematical Statistics},
title = {Structural properties and convergence results for contours of sample statistical depth functions},
volume = {28},
year = {2000}

}

@article{masse2004,
author = {Massé, Jean-Claude},
fjournal = {Bernoulli},
journal = {Bernoulli},
month = {06},
number = {3},
pages = {397--419},
publisher = {Bernoulli Society for Mathematical Statistics and Probability},
title = {Asymptotics for the Tukey depth process, with an application to a multivariate trimmed mean},
volume = {10},
year = {2004}
}

@article{Nolan92,
author = {Nolan, Deborah},
year = {1992},
month = {08},
pages = {157-169},
title = {Asymptotics for multivariate trimming},
volume = {42},
journal = {Stochastic Processes and their Applications},
doi = {10.1016/0304-4149(92)90032-L}
}

@article{Koshevoy97,
author = {Koshevoy, Gleb and Mosler, Karl},
fjournal = {The Annals of Statistics},
journal = {Ann. Statist.},
month = {10},
number = {5},
pages = {1998--2017},
publisher = {The Institute of Mathematical Statistics},
title = {Zonoid trimming for multivariate distributions},
volume = {25},
year = {1997}
}

@article{Dyckerhoff16,
author = {Dyckerhoff, Rainer},
year = {2016},
month = {11},
pages = {},
title = {Convergence of depths and depth-trimmed regions}
}

@article{Chen09, 
author={ Chen, Y. and Dang, X. and Peng, H. and Bart, H. L. J.}, 
journal={IEEE Transactions on Pattern Analysis and Machine Intelligence}, 
title={Outlier Detection with the Kernelized Spatial Depth Function}, 
year={2009}, 
volume={31}, 
number={2}, 
pages={288-305}, 
ISSN={0162-8828}, 
month={Feb}
}

@Article{Lange2014,
author="Lange, Tatjana
and Mosler, Karl
and Mozharovskyi, Pavlo",
title="Fast nonparametric classification based on data depth",
journal="Statistical Papers",
year="2014",
month="Feb",
day="01",
volume="55",
number="1",
pages="49--69",
issn="1613-9798"
}

@book{Anuj16,
  title={Functional and Shape Data Analysis},
  author={Anuj Srivastava and Eric P Klassen},
  isbn={978-1-4939-4018-9},
  year={2016},
  pages={83},
  publisher={Springer-Verlag New York}
}

\end{document}